\newcommand{\nc}{\newcommand}
\nc{\beq}{\begin{equation}}   \nc{\eeq}{\end{equation}}
\nc{\bea}{\begin{eqnarray}}   \nc{\eea}{\end{eqnarray}}
\nc{\baa}{\begin{array}}      \nc{\eaa}{\end{array}}
\nc{\bit}{\begin{itemize}}    \nc{\eit}{\end{itemize}}
\nc{\ben}{\begin{enumerate}}  \nc{\een}{\end{enumerate}}
\nc{\bce}{\begin{center}}     \nc{\ece}{\end{center}}
\def\beqa{\begin{eqnarray}}
\def\eeqa{\end{eqnarray}}

\def\half{{1\over 2}}

\def\lsim{\mathrel{\raise.3ex\hbox{$<$\kern-.75em\lower1ex\hbox{$\sim$}}}}
\def\gsim{\mathrel{\raise.3ex\hbox{$>$\kern-.75em\lower1ex\hbox{$\sim$}}}}

\def\ie{{\it i.e. }}

\def\be{\beq}
\def\ee{\eeq}
\def\to{\rightarrow}

\def\PHYSICA #1 #2 #3 {{\sl Physica}~{\bf#1} (#3) #2}
\def\MPL #1 #2 #3 {{\sl Mod.~Phys.~Lett.}~{\bf#1} (#3) #2}
\def\NPB #1 #2 #3 {{\sl Nucl.~Phys.}~{\bf #1} (#3) #2}
\def\NPBPS #1 #2 #3 {{\sl Nucl.~Phys.~B~(Proc. Suppl.)}~{\bf #1} (#3) #2}
\def\PLB #1 #2 #3 {{\sl Phys.~Lett.}~{\bf #1} (#3) #2}
\def\PR #1 #2 #3 {{\sl Phys.~Rep.}~{\bf#1} (#3) #2}
\def\PRD #1 #2 #3 {{\sl Phys.~Rev.}~{\bf #1} (#3) #2}
\def\PRL #1 #2 #3 {{\sl Phys.~Rev.~Lett.}~{\bf#1} (#3) #2}
\def\RMP #1 #2 #3 {{\sl Rev.~Mod.~Phys.}~{\bf#1} (#3) #2}
\def\ZPC #1 #2 #3 {{\sl Z.~Phys.}~{\bf #1} (#3) #2}
\def\IJMP #1 #2 #3 {{\sl Int.~J.~Mod.~Phys.}~{\bf#1} (#3) #2}

\documentstyle[preprint,epsfig,epsf,aps,floats]{revtex}
\begin{document}  
\newlength{\captsize} \let\captsize=\small 
\newlength{\captwidth}                     

\tightenlines

\title{Quest for the dynamics of $\nu_\mu \to \nu_\tau$ conversion}

\preprint{\vbox{\hbox{IFUSP-DFN/056-2000}}}

\author{A.\ M.\ Gago$^{1,2}$~\thanks{Email address: agago@charme.if.usp.br},
E.\ M.\ Santos$^{1}$~\thanks{Email address: emoura@charme.if.usp.br},
W.\ J.\ C.\ Teves$^{1}$~\thanks{Email address: teves@charme.if.usp.br} and 
R.\ Zukanovich Funchal$^{1}$~\thanks{Email address: zukanov@charme.if.usp.br}\\}

\vspace{1.cm}

\address{\sl 
$^1$ Instituto de F\'\i sica, Universidade de S\~ao Paulo,
    C.\ P.\ 66.318, 05315-970\\  S\~ao Paulo, Brazil\\
$^2$ Secci\'on F\'{\i}sica, Departamento de Ciencias,
    Pontificia Universidad Cat\'{o}lica del Per\'{u}, Apartado 1761\\
    Lima, Per\'{u}\\
\vglue -0.2cm
} 
   
\maketitle
\vspace{.5cm}
\hfuzz=25pt
\begin{abstract}
\noindent 
We perform a quantitative analysis of the capability of K2K, 
MINOS, OPERA and a neutrino factory in a muon collider 
to discriminate  the standard mass induced vacuum oscillation from the 
pure decoherence solution to the atmospheric neutrino problem and thereby 
contribute to unravel the dynamics that governs the 
observed $\nu_\mu$ disappearance.
\end{abstract}
\pacs{PACS numbers: }

\section{Introduction}
\label{sec:intro}

Two years ago, the Super-Kamiokande (SuperK) atmospheric neutrino data 
astonished the world giving the first compelling evidence in favor of 
$\nu_\mu \to \nu_\tau$ oscillation~\cite{sk98}. This incredible result 
has since been confirmed by other atmospheric neutrino 
experiments~\cite{soudan2,macro}, as well as by the preliminary 
K2K $\nu_\mu$ disappearance experiment result~\cite{k2k}, 
making unquestionable the fact that neutrinos suffer flavor 
conversion.

Naively, one may think that proves neutrinos have non-zero mass and that   
the next challenge for experimentalists is simply to determine the 
neutrino mass squared differences and the texture of the neutrino 
mixing matrix. Indeed, if the dynamics of neutrino flavor change is mass 
induced in the standard way~\cite{mns}, this is obviously the next logical 
step. Unfortunately, this is not an established fact. 

Although the atmospheric neutrino data collected up to now allow one to 
definite exclude some energy dependencies for 
the $\nu_\mu \to \nu_\tau$ conversion probability~\cite{lisi-vep}, some 
interesting possibilities, such as neutrino decay~\cite{bpl} and pure quantum 
decoherence~\cite{lisi} are capable of explaining the data comparably well 
to the standard mass induced oscillation mechanism.  
This in spite of the fact 
that the dynamics behind neutrino decay and pure 
decoherence give rise to a $\nu_\mu$ survival probability monotonically 
decreasing with the neutrino energy while the mass induced mechanism leads 
to a harmonic probability of oscillation. We therefore believe that another 
important experimental task should be to unravel the nature of the 
flavor changing  mechanism. 
We have to point out that the neutrino decay scenario is also mass induced, 
but throughout this paper when we allude to the mass induced mechanism, 
we will be referring to the standard neutrino flavor oscillation scenario.

Many, if not all, of the proposed future neutrino long baseline experiments 
were designed to measure $\nu_\mu \to \nu_\tau$ oscillations in order to 
pin down the oscillation parameters having in mind the standard mass induced 
oscillation mechanism. It is important to verify their real capabilities to 
discriminate among different flavor changing dynamics. 

The purpose of this paper is to investigate to what extent K2K and the 
next generation neutrino oscillation experiments will be able to 
discriminate the mass induced $\nu_\mu \to \nu_\tau$ oscillation 
solution~\cite{sk-atm} to the atmospheric neutrino problem (ANP),  
from  the recently proposed pure decoherence one~\cite{lisi}.
We do not investigate in this paper the decay 
mechanism, since it implies in the existence of sterile neutrinos, giving 
rise to a richer phenomenology, this study will be reported 
elsewhere~\cite{gptz}. 

The outline of the paper is as follows.
In Sec.\ \ref{sec:sec2}, we briefly review  the mass induced and the pure 
decoherence  mechanisms of flavor conversion.
In Sec.\ \ref{sec:sec3}, we define the statistical significance tests we 
will use in order to quantify the separation between the two ANP solutions. 
In Sec.\ \ref{sec:sec4}, we discuss the power of discrimination of 
the mass induced oscillation solution to the ANP from the 
dissipative one at K2K~\cite{k2kflux}, MINOS~\cite{minos}, OPERA~\cite{opera} 
and a possible neutrino factory in a muon collider~\cite{geer}.
Finally, in Sec.\ \ref{sec:sec5}, we present our conclusions.

\section{Review of the Formalism} 
\label{sec:sec2}

The time evolution of neutrinos created at a given flavor $\nu_\mu$ 
by  weak interactions, as of any quantum state,  can be described 
using  the density matrix formalism by the Liouville equation~\cite{fb}.
If we add an extra term $L[\rho_{\mu}]$ to the Liouville equation,  
quantum states can develop dissipation and irreversibility~\cite{fb,ellis1}. 
The generalized Liouville equation for $\rho_{\mu}(t)$ can then be written 
as~\cite{fb}

\begin{equation}
\displaystyle \frac{d \rho_{\mu}(t)}{d t}= -\imath [H,\rho_{\mu}(t)] +L[\rho_{\mu}(t)], 
\label{evol}
\end{equation} 
where the effective hamiltonian $H$ is, in vacuum, given by 
\begin{equation}
H= \left[ \begin{array}{cc}
 \Delta & 0 \\ 
0 & -\Delta 
\end{array} 
\right], 
\label{ham}
\end{equation} 
where $\Delta = (m^2_2-m^2_1)/4E_\nu$, we have already considered 
ultra-relativistic neutrinos of energy $E_\nu$ and the irrelevant global 
phase has been subtracted out. We assume here oscillation only between 
$\nu_\mu$ and $\nu_\tau$ in a two generation scheme.

The most general parametrization for $L[\rho]$ contains six real 
parameters which are not independent if one assumes the complete positivity 
condition~\cite{fb}. In one of the simplest situation, which in fact 
physically arises  when the weak coupling limit condition is 
satisfied, only one of the parameter, $\gamma$, has to be 
considered. In this limit, Eq.\ (\ref{evol}) can be solved to calculate 
the~\cite{fb} 

\begin{equation}
P(\nu_\mu \to \nu_\tau) = \frac{1}{2} \sin^2 2 \theta \; [1- e^{-2 \gamma L} \cos (2 \Delta L)],
\label{proba}
\end{equation} 
the probability of finding the neutrino produced in the flavor 
state $\nu_\mu$ in the flavor state $\nu_\tau$ after traveling a 
distance $L$ under the influence of quantum dissipation driven by the 
parameter $\gamma$.  

When $\gamma=0$, we get the usual mass induced oscillation (MIO) probability 
in two generation. The survival probability, in this case, is the 
standard one 
\begin{equation}
P(\nu_\mu \to \nu_\tau) = \frac{1}{2} \sin^2 2 \theta \; [1- \cos (2 \Delta L)].
\label{pmio}
\end{equation} 

On the other hand, if neutrinos are massless or degenerate ($\Delta m^2 =0$) 
and weak interaction eigenstates are equal to mass eigenstates, even though 
standard oscillations cannot occur, flavor conversion can still take place 
through the pure decoherence mechanism (PDM)~\cite{gstz}. 
Explicitly, the neutrino flavor change probability, in the simplest case 
where a single decoherence parameter is considered, becomes

\begin{equation}
P(\nu_\mu \to \nu_\tau) = \frac{1}{2} [1- e^{-2 \gamma L}].
\label{ppdm}
\end{equation} 

We will assume here for PDM that $\gamma= \gamma_0 (E_\nu/\text{GeV})^{-1}$, 
where $\gamma_0$ is a constant given in GeV.
This {\em Ansatz} may be motivated by the assumption that the exponent in 
Eq.~(\ref{ppdm}) behaves like a scalar under Lorentz transformations~\cite{lisi}.

Note that here we will be studying both MIO and PDM in a two generation 
framework since this is enough to explain well the atmospheric neutrino data. 
Notwithstanding, one may wonder about the contribution from electron 
neutrinos. If PDM really takes place in nature, it should in fact involve 
all three neutrino flavors. We can have some insight 
on what can happen in this case. 
From the results of Refs.\ \cite{gstz,kps} on the limits on PDM from 
SN 1987A data, and with the {\em Ansatz} above, we expect the decoherence 
parameter that accompany the $\nu_\mu \to \nu_e$ conversion to be smaller 
than $10^{-39}$ GeV. This means that for the energies and distances of the 
long baseline experiments we will study here, this effect should be 
completely negligible. Nevertheless, a complete study of the PDM with 
three neutrino generations should in fact be performed to confirm this 
assumption.

\section{Statistical Significance Test}
\label{sec:sec3}

In order to define the capability of an experiment to discriminate 
mass induced $\nu_\mu \to \nu_\tau$ oscillation  from pure decoherence, 
we define the number of standard deviations of 
separation between MIO and PDM as $n_\sigma = \sqrt{\chi^2}$ where   

\bea
\chi^2(\gamma_0,\sin^2 2\theta, \Delta m^2) &=& 
2 [N^{\text{PDM}}(\gamma_0)-N^{\text{MIO}}(\sin^2 2\theta, \Delta m^2)] \nonumber \\
&+& 2 N^{\text{MIO}}(\sin^2 2\theta, \Delta m^2) \; \ln \left( \frac{N^{\text{MIO}}(\sin^2 2\theta, \Delta m^2)}{N^{\text{PDM}}(\gamma_0)}\right),
\label{conf-l}
\eea
is the confidence level according to the procedure proposed by the 
Particle Data Book~\cite{pdg00}. Here, $N^{\text{PDM}}(\gamma_0)$ is the total 
number of events theoretically expected if PDM is the solution to the ANP and 
$N^{\text{MIO}}(\sin^2 2\theta, \Delta m^2)$ is the total number of 
events that can be observed by the experiment as a function of the 
two parameters involved in the MIO mechanism.

For each experiment we have studied, we have computed two different types of 
contour level curves. 

First, we fix $N^{\text{PDM}}=N^{\text{PDM}}(\gamma_0^{\text{best}})$,
 at the number corresponding to $\gamma_0 =\gamma_0^{\text{best}}
= 0.6 \times 10^{-21}$ GeV, the best-fit point of the PDM 
solution to the ANP~\cite{lisi} and vary the MIO parameters in the 
interval $1 \times 10^{-3} \text{ eV}^2 \leq \Delta m^2 \leq 2 \times 10^{-2}$ eV$^2$ and $0.8 \leq \sin^2 2\theta \leq 1$ consistent with the atmospheric 
neutrino data. In this way, we obtain curves of fixed $n_\sigma$ in the plane 
$\sin^2 2 \theta \times \Delta m^2$.

Second, we fix $\sin^2 2\theta=1$ and vary the PDM parameter 
$\gamma_0$ in the interval $0.25 \times 10^{-21} \text{ GeV} \leq \gamma_0 \leq 1.1 \times 10^{-21} \text{ GeV}$, as well as 
 the MIO parameter $\Delta m^2$  in the interval 
$1 \times 10^{-3} \text{ eV}^2 \leq \Delta m^2 \leq 2 \times 10^{-2}$ eV$^2$. 
The upper limit of the range in $\gamma_0$ is the one allowed by the 
CHORUS/NOMAD data~\cite{lisi,gstz}, the lower limit was estimated  
by using $\gamma_0 \sim 2.54 \times 10^{-19} \times (\Delta m^2/\text{eV}^2)$ GeV, with 
$\Delta m^2 = 1.0 \times 10^{-3}  \text{ eV}^2$.
Like this, we can get  curves of fixed $n_\sigma$ in the plane 
$\gamma_0 \times \Delta m^2$.
This allows us to extend our conclusions beyond the best-fit value of the PDM
solution to the ANP.

\section{PDM versus MIO}
\label{sec:sec4}

We have investigated the capability of K2K~\cite{k2kflux,ratio,ishida} and 
the next generation neutrino oscillation experiments MINOS~\cite{minos}, 
OPERA~\cite{opera,ngs} and a neutrino factory in a muon collider~\cite{geer}, 
to discriminate the PDM solution to the ANP with 
$\gamma_0 \sim 0.6 \times 10^{-21}$ GeV~\footnote{We remark that in our 
notation $2 \gamma_0$ corresponds to $\gamma_0$ of Ref.\ \cite{lisi}.},  
using the ansatz $\gamma = \gamma_0(E_\nu/\text{GeV})^{-1}$, given in 
Ref.~\cite{lisi}, from the traditional one due to $\nu_\mu \to \nu_\tau$ 
MIO in vacuum with $\Delta m^2 \sim (1.1-7.8) \times 10^{-3}$ eV$^2$ 
and $\sin^2 2 \theta \gsim 0.84$~\cite{sk-atm}. We recall that the 
best-fit point for the MIO solution to the ANP is at $(\sin^2 2\theta, 
\Delta m^2)=(1.0,3.0\times 10^{-3} \text{ eV}^2)$~\cite{sk-atm}.

We would like to point out that the decoherence solution to the ANP is open 
to two different readings: either it can be viewed as an effect of  pure 
decoherence  or as a combination of quantum decoherence 
plus vacuum oscillation driven by $\Delta m^2 \lsim 10^{-6}$ eV$^2$ 
with $\sin^2 2\theta \sim 1$. 
In the first case, there is a single free parameter ($\gamma_0$) and 
the flavor change probability is given by Eq.~(\ref{ppdm}), 
in the second, there are two free parameters ($\gamma_0$ and $\sin^2 2\theta$) 
since the probability will be given 
by Eq.~(\ref{proba}) with $\cos(2 \Delta L) \rightarrow 1$.

We now present and discuss the results of our study.

\subsection{K2K}
\label{k2ksec}
In Ref.\ \cite{lisi}, K2K was cited as a possible experiment to test 
the novel decoherence solution to the ANP. 
This possibility would be very appealing for K2K is an experiment 
which is currently taking data.

In order to verify this, we have calculated the expected number of events 
in K2K for the goal of the experiment, \ie $10^{20}$ protons on 
target (POT)~\cite{k2k}, for three  hypotheses: no flavor conversion, 
mass induced $\nu_\mu \to \nu_\tau$  oscillation with parameters 
consistent with SuperK atmospheric neutrino results~\cite{sk-atm} 
and the pure decoherence solution to the ANP~\cite{lisi}. 

K2K is a $\nu_\mu \to \nu_\mu$ disappearance experiment, 
where the muon neutrinos that have an average energy of 1.4 GeV are 
produced by the KEK accelerator, first measured after traveling 300 m 
by the Near Detector, which is a 1 kton water Cherenkov detector, 
and finally measured by the Far Detector (FD),  the SuperK   
22.5 kton water Cherenkov detector localized at 250 km from the target. 
This experiment, which has started taking data last year and has currently 
accumulated $2.29 \times 10^{19}$ POT, seems to be confirming 
$\nu_\mu \to \nu_\mu$ disappearance as expected by the atmospheric neutrino 
results~\cite{sk-atm}.

The expected number of events in the FD can be computed as 
follows

\be
N_{\text{FD}} = R \, N_{\text{FD}}^{\text{theo}},  
\label{k2kexp}
\ee
where $R$ is the ratio between the number of observed over the number of 
expected events in the Near  Detector,  we have used the fact that 
$R \sim 0.84$ from Ref.~\cite{ratio}, and  $N_{\text{FD}}^{\text{theo}}$  
is the theoretical expectation that can be calculated as 

\be
N_{\text{FD}}^{\text{theo}}
=  \eta \; n_{\text{FD}} \int \Phi_{\text{FD}}(E) \sigma(E) P(\nu_\mu \to \nu_\mu) dE ,
\label{nsk}
\ee
where $E$ is the neutrino energy, $\Phi_{\text{FD}}(E)$ is the $\nu_\mu$ flux 
distribution at the Far Detector, $\sigma(E)$ is the total neutrino 
interaction cross section taken from Ref.\ \cite{x-sec-k2k} 
and $n_{\text{FD}}$ is the number of active targets 
in the FD. Also, we have introduced a normalization factor $\eta$ 
which was fixed to 0.65 in order to get the same expected number of 
$\nu_\mu$ events as K2K for null oscillation. 
The shape of $\Phi_{\text{FD}}(E)$, 
was taken from  Ref.\ \cite{k2kflux}, but the total flux has been 
renormalized to account for the number quoted in Table 1 of Ref.\ \cite{ratio}.
The survival probability $P(\nu_\mu \to \nu_\mu) = 1- P(\nu_\mu \to \nu_\tau)$
in two generations, with $P(\nu_\mu \to \nu_\tau)$ either equal to zero 
(for no $\nu_\mu \to \nu_\tau$ conversion), to the usual two generation MIO 
probability (Eq.~(\ref{pmio})) or the PDM flavor conversion 
probability (Eq.~(\ref{ppdm})). 
 
We first perform the calculation of the total number of expected events, 
$N_{\text{FD}}$, in the absence of any flavor change 
and for oscillation with $\sin^2 2 \theta=1$ and 
$\Delta m^2 = 3\times 10^{-3}$ eV$^2$, 
$5\times 10^{-3}$ eV$^2$ and $7 \times 10^{-3}$ eV$^2$ for 
2.29 $\times 10^{19}$ POT.
These results, which agree quite well with the K2K estimations presented in 
Ref.\ \cite{k2k},  are summarized in Table \ref{tab1}. 
Thus we are confident that our numbers are reasonable and we can proceed to 
estimate the total number of events expected in the FD when K2K 
reaches $10^{20}$ POT for vacuum oscillation and decoherence. 
These numbers are also reported in Table~\ref{tab1}.  From this, we can see
that the mass induced oscillation and the decoherence effect at their 
best-fit values imply, for the goal of the K2K experiment,
values for the total number of $\nu_\mu$ events which are statistically 
compatible, hence the two solutions will be indistinct at K2K.    

\begin{figure}
\vglue -3.0cm
\centerline{
\epsfig{file=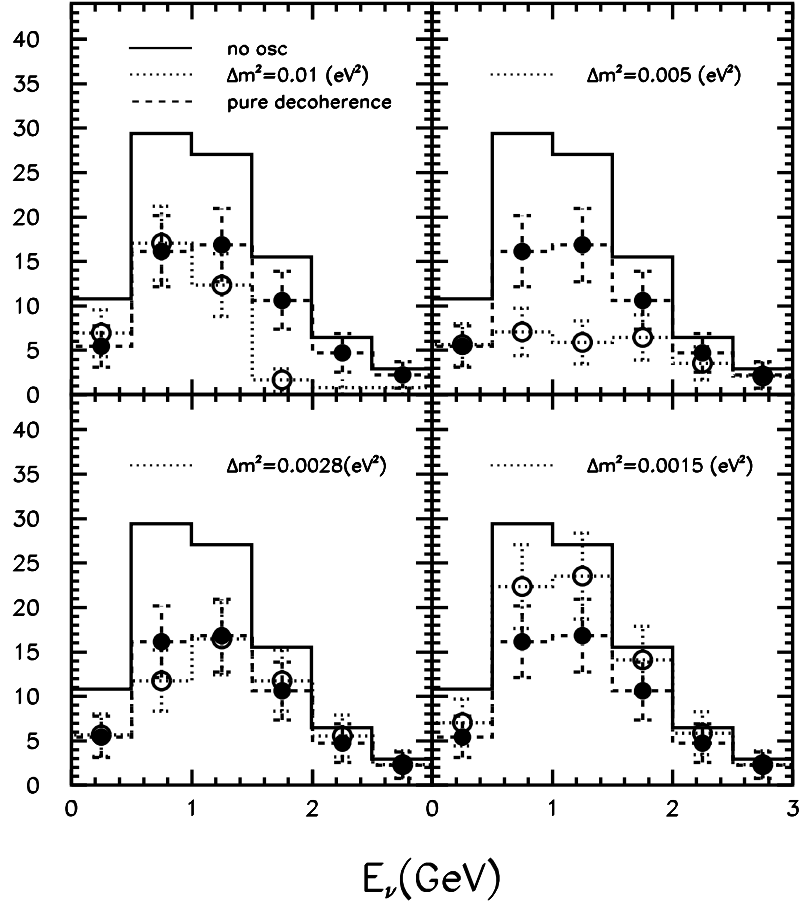,height=20.8cm,width=15cm}}
\vglue -4.5cm
\caption{Spectral distortion expected for one ring $\mu$-like events at 
K2K assuming  $\nu_\mu \to \nu_\tau$ flavor conversion with  
$\sin^2 2 \theta=1$ and different values of $\Delta m^2$ (MIO) 
and $\gamma_0=0.6\times 10^{-21}$ GeV (PDM).
The bars represent the statistical error at each bin.
The calculation was done for a total of $10^{20}$ POT.}
\label{fig1a}
\vglue -0.05cm
\end{figure}

One may wonder about the energy distribution of the K2K events, which, 
in principle, could be used to discriminate the solutions.
We show, in Fig.~\ref{fig1a}, the Monte Carlo simulated reconstructed 
neutrino spectrum for one ring $\mu$-like events at SuperK taken from 
Ref.~\cite{ishida} for $\Delta m^2=0.0015 \text{ eV}^2, 
0.0028 \text{ eV}^2,0.005 \text{ eV}^2 \text{ and } 0.01 \text{ eV}^2$ and 
$\sin^2 2\theta=1$, as well as our estimation of the distortion expected for 
the best-fit point of the PDM solution to the ANP.
The latter was done simply by multiplying the bin content for no oscillation 
by the average PDM survival probability in the bin.

From Fig.~\ref{fig1a}, we see that if one takes into account only the 
statistic error, completely disregarding the systematic one, the curves 
are already virtually indistinguishable so that one unfortunately can hardly 
hope to discriminate between these two solutions with the K2K data. 
It is important to remark that the actual spectrum is currently under study,  
so what we present here can be viewed as a tendency which indicates that, 
even if one compares only the shape for the best-fit values of the two 
solutions, this type of discrimination will be very difficult in K2K. 

In Figs.~\ref{fig1b} and \ref{fig1c}, we show the result of the 
statistical significance tests for K2K,  
as proposed in Sec. \ref{sec:sec3}. 
These curves were calculated with the total number of events $N_{\text{FD}}$ 
and, although this is not directly related to what is plotted in 
Fig.~\ref{fig1a}, it takes us basically to the same conclusion, \ie for 
data compatible with 
$2.2\times 10^{-3} \text{ eV}^2 \leq \Delta m^2 \leq 4.0 \times 10^{-3} 
\text{ eV}^2$ the maximal separation between the two solutions cannot 
exceed $\sim$ 2 $\sigma$ if $\gamma_0=\gamma_0^{\text{best}}$. 
We see in Fig.~\ref{fig1c} that  
for data consistent with
 $2.2\times 10^{-3} \text{ eV}^2 \leq \Delta m^2 \leq 3.5 \times 10^{-3} 
\text{ eV}^2$  the separation is always less 
than 3 $\sigma$, $\forall \gamma_0$.  A 
separation of 5 or more $\sigma$ can only be achieved if the data is 
compatible with $\Delta m^2 \gsim 7 \times 10^{-3} \text{ eV}^2$.
We also see that the point which correspond to the best-fit of the MIO and 
the PDM solutions is inside the $n_\sigma \lsim 1$ region.  
Therefore, we conclude that it will be rather difficult to disentangle the 
two ANP solutions before the arrival of the next generation neutrino 
experiments. 

\subsection{MINOS}
\label{minos}

The MINOS experiment~\cite{minos} is  part of the Fermilab NuMI Project.
The neutrinos which constitute the MINOS beam will be  the result of the  
decay of pions and kaons that will be produced by the 120 GeV proton high 
intensity beam extracted from the Fermilab Main Injector. There will be two 
MINOS detectors, one located at Fermilab (the near detector) and  another 
located in the Soudan mine in Minnesota, about 732 km away (the far detector 
of 5.4 kton). 

According to Ref.~\cite{minostec}, MINOS will be able to measure 
independently the rates and the energy spectra for muonless ($0 \mu$) and 
single muon ($1 \mu$) events, which are related to the neutral 
current (nc) and charged current (cc) reactions.
Three different neutrino energy regions are 
possible: low ($E_\nu \sim 3$ GeV), medium ($E_\nu \sim 7$ GeV) and high
($E_\nu \sim 15$ GeV). For MINOS to operate as a $\nu_\tau$ appearance 
experiment either the high or medium energy beam is required, since one must 
be above the $\tau$ threshold of 3.1 GeV. 

We have studied here two different observables that can be measured in 
MINOS : the $1 \mu$-event energy spectrum and the $0 \mu$/$1 \mu$ 
event ratio.
The three different beam possibilities were investigated. 

The expected number of $1 \mu$-events in MINOS, 
$dN_{1 \mu}$, can be calculated using~\cite{numi}

\be
\displaystyle \frac{dN_{1\mu}}{dE_\nu}(E_\nu) = 
\displaystyle \frac{dN_{\text{cc}}}{dE_\nu}(E_\nu)  
\left \{ [1-P({\nu_\mu \rightarrow \nu_\tau})] + 
P({\nu_\mu \rightarrow \nu_\tau}) \, \eta(E_\nu) \, \text{Br}(\tau \to \mu) 
\right \},
\label{oscmin}
\ee
where $\eta(E_\nu)=\sigma_{\nu_\tau-\text{cc}}(E_\nu)/\sigma_{\nu_\mu-\text{cc}}(E_\nu)$, is 
the ratio of the cc-cross section for $\nu_\tau$ over the 
cc-cross section for $\nu_\mu$, $\text{Br}(\tau \to \mu)$ is the  branching 
ratio of the tau leptonic decay to muon (18\%),
$dN_{\text{cc}}/dE_\nu$ is the energy 
spectrum for $\nu_\mu$ cc-events in the MINOS far detector in the case 
of no flavor change~\cite{minostec} and 
$P({\nu_\mu \rightarrow \nu_\tau})$ is the probability of 
$\nu_\mu \to \nu_\tau$ conversion.  

There are two different ways a muon can be produced; either by the 
surviving $\nu_\mu$ which interact with the detector (first term of 
Eq.~(\ref{oscmin})) or by the contribution from  
taus  generated by $\nu_\tau$ interactions in the detector, after  
$\nu_\mu \to \nu_\tau$ conversion, followed by  
$\tau \rightarrow \nu_\mu \nu_\tau \mu$ decay 
(second term of Eq.~(\ref{oscmin})). 
In both cases, the events must trigger the detector and  be identified 
as muons to count as $1 \mu$-events. Here, we have not considered the 
possible contamination from neutral current events and  the trigger and 
identification efficiencies were supposed to be 100\%. 

In Fig.~\ref{fig2a}, we present the $1 \mu$-event energy 
distribution expected at MINOS for no flavor change and the best-fit 
parameters of the MIO and of the PDM solutions to the ANP for 10 kton year 
($\sim$ 2 years of running). 
We can observe that if the final choice of beam is the  low energy 
configuration, MINOS will be in the same footing of K2K, meaning that 
discrimination between solutions will be extremely disfavored due to low 
statistics. In the event of the choice fall on the medium or the high 
energy beam, discrimination will become more likely due to higher statistics.

We show in Fig.~\ref{fig2b} the result of the statistical 
significance test in the $\Delta m^2 \times \sin^2 2\theta$ plane,
for the three beam setups. 
We see that for $\Delta m^2= 3 \times 10^{-3}$ eV$^2$, 
PDM and MIO are  separated by less than 2 $\sigma$ (low), more than 
3 $\sigma$ (medium) and more than 5 $\sigma$ (high), $\forall \sin^2 2\theta$ 
in the range compatible with SuperK atmospheric data, confirming our 
previous  conclusions on Fig. \ref{fig2a}.

We can further point out that, considering the 99 \% C.L. 
$\sin^2 2\theta \times \Delta m^2$ region allowed by the SuperK  
data~\cite{sk-atm}, the two solutions will be indistinguishable 
within 3 $\sigma$ for $ 1.7 \times 10^{-3} \text{ eV}^2 \lsim \Delta m^2 
\lsim 5.3 \times 10^{-3} \text{ eV}^2$ in the low energy configuration,
for $ 3.0 \times 10^{-3} \text{ eV}^2 \lsim \Delta m^2 
\lsim 4.3 \times 10^{-3} \text{ eV}^2$ in the medium energy configuration and
for $ 4.2 \times 10^{-3} \text{ eV}^2 \lsim \Delta m^2 
\lsim 6.2 \times 10^{-3} \text{ eV}^2$ in the high energy configuration.

In Fig.~\ref{fig2c}, we show the statistical significance test in the 
plane $\gamma_0 \times \Delta m^2$ for the three energy possibilities. 
The point which correspond to the best-fit of the MIO and 
the PDM solutions is inside the $n_\sigma \lsim 1$ region for low, 
the  $n_\sigma \lsim 2$ region for medium and only for the high energy 
case it falls into the  $n_\sigma \gsim 5$ region.  
Here it is again demonstrated that the medium and high energy setups 
are preferable to the low one if the issue is to discriminate the 
two mechanisms. 

We have also computed the ratio $R_{{0 \mu}/{1 \mu}}$ 
that should be expected in MINOS as a function of the free parameters of the 
flavor changing hypothesis. This ratio has the advantage that it does not 
require the understanding of the relative fluxes at the near and far 
detectors, it is also quite sensitive to neutrino flavor conversion since 
when they occur not only $1 \mu$-events are depleted but 
$0 \mu$-events are enhanced.
This ratio can be written as~\cite{numi}

\be
R_{{0 \mu}/{1 \mu}} = 
\frac{\int dN_{0 \mu}(E_\nu)} 
{\int dN_{1 \mu}(E_\nu)}.
\label{ratio1}
\ee 

The number of expected $0 \mu$-events can be calculated as 

\be
\displaystyle \frac{dN_{0 \mu}}{dE_\nu} (E_\nu)
=  [ 
\displaystyle \frac{dN_{\text{nc}}}{dE_\nu}(E_\nu)+  \displaystyle \frac{dN_{\text{cc}}}{dE_\nu}(E_\nu)
P({\nu_\mu \rightarrow \nu_\tau}) \, \eta(E_\nu) \, (1-\text{Br}(\tau \to \mu))
],
\label{oscminos}
\ee
where we again supposed no contamination and the trigger and identification 
efficiencies to be 100\%. 
We can  infer $dN_{\text{nc}}/dE_\nu$, 
the expected energy spectrum for $\nu_\mu$ nc-events in the MINOS far 
detector in the case of no flavor change, using the approximation

\be
\displaystyle \frac{dN_{\text{nc}}}{dE_\nu}(E_\nu) \sim 
\displaystyle \frac{dN_{\text{cc}}}{dE_\nu}(E_\nu)
\frac{\sigma_{\nu_\mu-\text{nc}}(E_\nu)}
{\sigma_{\nu_\mu-\text{cc}}(E_\nu)}.
\ee

The cross sections $\sigma_{\nu_\mu-\text{cc}}$, 
$\sigma_{\nu_\tau-\text{cc}}$ were taken from Ref.\ \cite{x-sec} 
and $\sigma_{\nu_\mu-\text{nc}}$ from Ref. \cite{x-sec-k2k}.

In Table~\ref{tab2}, we show our estimation of this ratio for no 
$\nu_\mu \to \nu_\tau$ conversion, mass induced oscillation for  
several values of $\Delta m^2$ and PDM with 
$\gamma_0=\gamma_0^{\text{best}}$ for the 
medium and high energy beam. For the low energy beam this test is 
ineffective since, in this case, MINOS will work essentially 
like a $\nu_\mu$ disappearance experiment. 
For the medium and high energy beams, the ratio predicted for the best-fit 
values of the parameters of the two solutions seem to be well separated. 
Nevertheless one has to take these results carefully, in a real experimental 
situation,  experimental efficiencies, event contamination, resolution and
systematic errors can substancially affect $0 \mu$-event observables.

\subsection{CERN-to-Gran Sasso Neutrino Oscillation Experiments}
\label{gsexp}
The new CERN neutrino beam to Gran Sasso is a facility 
that will direct a $\nu_\mu$ beam to the Gran Sasso Laboratory in Italy, 
at 732 km from CERN. Such a beam together with the massive detectors 
ICANOE/OPERA~\cite{ngs} in Gran Sasso will constitute a powerful tool for 
long baseline neutrino oscillation searches. The number of protons on target 
is expected to be 4.5 $\times 10^{19}$ per year, the $\nu_\mu$ beam will have 
an average energy of 17 GeV, the fractions  $\nu_e/\nu_\mu$, $\nu_{\bar{\mu}}/\nu_\mu$  
and $\nu_\tau/\nu_\mu$  in the beam are expected to be as low as 
0.8\%, 2\% and $10^{-7}$, respectively~\cite{icanoe}.
The number of $\nu_\mu$ charged current events per protons on target 
and kton,  without neutrino oscillation, is calculated to be 
$4.7 \times 10^{-17}$, for $1$ GeV $\le E_\nu \le 30$ GeV~\cite{icanoe}.
These experiments are supposed to start taking data around 2005 and may be 
used to try to distinguish the two ANP solutions.

We have investigated the capability of these detectors, in particular OPERA,
working in the $\nu_\mu \to \nu_\tau$ appearance mode, to elucidate which 
is the correct solution to the ANP. 
We have obtained the expected number of $\nu_\tau$ events, $N_\tau$, that will 
be measured by OPERA, considering a pure $\nu_{\mu}$ beam, 
using the following expression 
\begin{equation}
N_\tau= A \int \phi_{\nu_{\mu}}(E) \times P(\nu_{\mu} \to 
\nu_{\tau}) \times \sum_{i=l,h} \text{Br}(\tau \to i) \times \sigma_{\nu_{\tau}-\text{cc}}(E) \times \epsilon_i(E) \; dE,
\label{nutau}
\end{equation}
where $\phi_{\nu_{\mu}}$ is the flux of $\nu_\mu$ at the Gran Sasso 
detector and  $\sigma_{\nu_{\tau}-\text{cc}}$ is the charged current 
cross section for $\nu_{\tau}$, both taken from Ref.~\cite{x-sec}.  
The number of active targets $A$ can be calculated as 
$A = C_n \times M_{d} \times  N_{A} \times 10 ^{9}\times N_{p} \times N_y$ 
where $M_{d}$ is the detector mass in kton,  $N_{A} \times 10^{9}$ is 
the number of nucleons per kton (where $N_{A}=6.02 \times 10^{23}$ is the 
Avogadro's number),  $N_y$ is the number of years of data taking and $N_{p}$ 
is the number of protons on target per year and $C_n=0.879$ is normalization  
constant that we need to introduce in order to be able to reproduce  
the numbers presented in Table 27 of Ref.~\cite{opera}.
We have assumed that $\nu_\tau$ identification 
will be accomplished through its one-prong decays into leptons ($l$) and 
hadrons ($h$).
We have used the overall value $\sum_{i=l,h} \text{Br}(\tau \to i) \times \epsilon_i(E) = 8.7$~\%, admitting a total mass of 2 kton for 5 years of 
exposure, according to Ref.~\cite{opera}.
The probability $P( \nu_{\mu} \to \nu_{\tau})$ is either supposed 
to be equal to zero (for no flavor transformation), to the MIO probability 
(Eq.\ (\ref{pmio})) or to the PDM probability (Eq.\ (\ref{ppdm})).

In Table~\ref{tab3}, we show the number of $\nu_\tau$ events we have 
calculated, according to Eq.~(\ref{nutau}), for 5 year exposure, 
assuming  $\sin^2 2\theta=1$ and four different values of $\Delta m^2$ and the 
best-fit point of the decoherence solution to the ANP. 
We also quote in this table the total number of background events, which 
remains after all the kinematical cuts have been applied, normalized to an 
exposure of 5 years, taken from Ref.~\cite{opera}.  
We observe that for the decoherence effect the rate of tau events is 
substantially higher than that of most of the $\Delta m^2$ hypotheses 
and than the number of background events.

We show in Fig.~\ref{fig3a} the calculated energy spectrum for both flavor 
changing scenarios at OPERA for 5 years of exposure,  this was calculated
using Eq.\ (\ref{nutau}). In spite of the fact that in some cases we observe 
that the ratio between the number of $\nu_\tau$ events coming from PDM and 
MIO is higher than four, in practice, it seems to be very difficult to 
observe this difference due to low statistics.

In Fig.~\ref{fig3b}, we show the result of the first statistical significance 
test. From this we see that discrimination between PDM and MIO can become 
difficult if $\Delta m^2 \gsim 4.4\times 10^{-3}$ eV$^2$ ($n_\sigma \lsim 5$).

It is worth mentioning that if we vary the decoherence parameter in the range 
$0.25 \times 10^{-21} \text{ GeV} \lsim \gamma_0 \lsim  1.1 \times 10^{-21} \text{ GeV}$, the number of expected $N_\tau$ events also varies from 31 to 114.
This suggests that there might be some situations, even if the measured 
number of events is consistent with  $\Delta m^2 \lsim 4.4 \times 10^{-3}$ 
eV$^2$, where it might be difficult to disentangle the two ANP solutions 
 in  OPERA.
This becomes clearer in Fig.~\ref{fig3c}, where we see that for data 
compatible with $\Delta m^2 < 4.1 \times 10^{-3} \text{ eV}^2$ a separation 
of more than 3 $\sigma$  can be achieved for 
$\gamma_0 \gsim 4 \times 10^{-21}$ GeV, but data consistent with 
values of $\Delta m^2\sim 5.3 \times 10^{-3}$  eV$^2$ can only be separated 
from the PDM solution if $\gamma_0 \gsim 7 \times 10^{-21}$ GeV 
($n_\sigma \gsim 5$). 
 
\subsection{Neutrino Factory in Muon Collider}
\label{numumu}

Many authors~\cite{geer,deruj,bgw} have emphasized  the advantages of 
using the straight section of a high intensity muon storage ring to 
make a neutrino factory. The muons (anti-muons) accelerated to an 
energy $E_\mu$ ($E_{\bar{\mu}}$) constitute a pure source of 
both $\nu_\mu$ ($\bar \nu_\mu$) and $\bar \nu_e$ ($ \nu_e$) through their 
decay $\mu^- \to e^- \bar \nu_e \nu_\mu$ ($\mu^+ \to e^+  \nu_e \bar \nu_\mu$)
with well known initial flux and energy distribution.

In the many propositions for this type of long baseline neutrino factory one 
can find in the literature, the stored muon (anti-muon) energy $E_\mu$ 
($E_{\bar{\mu}}$)  ranges from 10 GeV to 250 GeV  and the neutrino beam is 
directed towards a faraway detector at a distance corresponding to an 
oscillation baseline $L$ varying from 730 km to 10\, 000 km.

Here we have explored the neutrino factory as a disappearance 
$\nu_\mu \to \nu_\mu$ experiment.  We will explicitly discuss the case when 
negative muons are stored in the ring, a similar calculation can be performed 
when positive muons are the ones which decay producing neutrinos.  
The relevant observable is the total number of $\mu$ that can be detected 
when $\mu$ is the produced charged lepton in the beam, \ie the number of 
``same sign muons'' that we will denote here by $N_{\mu}$. We define 
$N_{\mu}$, for unpolarized muons (see Appendix), as

\beqa
N_{\mu} & = & n_{\mu} M_d \frac{10^9 N_{\text{A}}}{m_\mu^2 \pi} \frac{E^3_\mu}{L^2} \int_{E_{\text{th}}/E_\mu}^{1} 
 h_0 (x) \epsilon_{\mu}(xE_\mu) \left \{ \frac{\sigma_{\nu_{\mu}-\text{cc}}(xE_\mu)}{E_\mu}  [1-P(\nu_\mu \to \nu_\tau)] \right. \nonumber \\ 
& + & \left.  \, \frac{\sigma_{\nu_{\tau}-\text{cc}}(xE_\mu)}{E_\mu} \, \text{Br}(\tau \to \mu) \, P(\nu_\mu \to \nu_\tau) \right \}  dx,
\label{evtn}
\eeqa
where $E_\mu$ is the muon source energy, $x=E_\nu/E_\mu$, $M_d$ is the 
detector mass in ktons, $n_{\mu}$ number of useful $\mu$ decays, 
$10^{9} N_{\text{A}}$ is the number of nucleons in a kton and $m_\mu$ is 
the mass of the muon.  The function $h_0$ contain the $\nu_\mu$ energy 
spectrum normalized to 1 explicitly given in the Appendix.
The charged current interaction cross sections per nucleon 
$\sigma_{{\nu_\mu}-\text{cc}}$ and $\sigma_{{\nu_\tau}-\text{cc}}$ can be 
found in Ref.\ \cite{x-sec}. 
This number has two contributions, one from the $\nu_\mu$ produced in the 
decay that survive and arrive at the detector interacting with it producing 
a final $\mu$, another from the $\nu_\tau$ that is produced by the flavor 
conversion mechanism and interact in the detector via charge current 
producing $\tau$, which subsequently decays to $\mu$ with a branching ratio 
$\text{Br}(\tau \to \mu)$. 
We have calculated the above integral from $E_{\text{th}} = 3$ GeV  to ensure 
good detection efficiency~\cite{flr}, so that we can consider the $\mu$  
detection efficiency $\epsilon_{\mu}$ to be 100 \%, independent of energy.

The muon beam is expected to have an average angular divergence of 
the ${\cal O}(0.1\;m_\mu/E_\mu)$. 
It was pointed out in Ref.\ \cite{bgw} that this 
effect is about 10 \% so we have multiplied our calculated number of events 
by 0.9 to account for this. 
The contribution of the background to the number of muons 
observed in the detector, that includes muons from charm decays produced by 
charged current and neutral current interactions in the detector, has 
been neglected on account that this would be a small global contribution.

Since the final configuration  of such a facility is still not defined,    
we have tried to estimate the optimum configuration in order to maximize the 
difference between mass induced $\nu_\mu \to \nu_\mu$ survival probability 
and pure decoherence. As observed in Ref.\ \cite{lisi}, at the best-fit 
point of the MIO and PDM solutions, the argument of the cosine, 
in the MIO case, and the argument of the exponential, in the PDM case, can 
be viewed to be approximately the same. Explicitly,
\be
 P(\nu_\mu \to \nu_\mu) \simeq \half [ 1+\cos (X)] \text{ (MIO) },
\ee

\be
 P(\nu_\mu \to \nu_\mu) \simeq \half [ 1+e^{-X}] \text{ (PDM) },
\ee
where 
\be
X = \beta \frac{L}{E_\nu},
\label{X}
\ee
with $\beta = 2.54 \;  \Delta m^2  \sim 1.0 \times 10^{19} \; \gamma_0 $, 
 $\Delta m^2$ given in eV$^2$, $\gamma_0$ in GeV and  $\beta$ in GeV/km. 
So that we can maximize the difference between the probabilities simply by 
finding the values of  $X$, which maximize the function

\be
F(X) = | \cos (X) - e^{-X} |.
\label{optl}
\ee

This means that one has to numerically solve for $X$ the 
transcendental equation  
\be
 e^{-X} - \sin(X) = 0,
\label{trans}
\ee
with the condition $\cos(X) > - e^{-X}$.

Once we find the spectrum of solutions ${X}$ we can apply Eq.\ (\ref{X}) to 
find the optimal distance $L_{\text{opt}}$ by choosing  a value for $E_\mu$ 
and fixing $E_\nu$ at the average value of the observable neutrino energy, 
\ie $ \langle E_\nu \rangle = 0.7 \; E_\mu$ \cite{bgrw}, with  $\beta = 6.1 \times 10^{-3}$ 
GeV/km, the best-fit value of Ref.\ \cite{lisi}.  
We have calculated the  maximal difference between the two survival  
probabilities for $E_\mu= 10 \text{ GeV}, 20 \text{ GeV}, 30$ GeV 
and 50 GeV as a function of $L$,  but 
only kept the cases where this difference reaches about 50 \%.  
This was obtained for $L_{\text{opt}}= $ 3096 km, 6192 km, 9289 km  
and $E_\mu= 10 \text{ GeV}$, 20 GeV and 30~GeV, respectively. 

Note that using this criteria, we are looking for a maximum in the absolute 
difference between the survival probabilities (${\cal{O}}(0.5)$), 
which is a very strong condition that we impose in order to detect and 
distinguish the signal of $\nu_\mu$ disappearance. Since this depends on 
the number of muon decays, which is big, this ensures that the signal will 
be quite sizable experimentally. We have seen before 
(see for instance Fig.\ref{fig3a}), that there are cases of $L/E$ where the
absolute difference in the survival probabilities is small while the  
ratio $P(\nu_\mu \to \nu_\tau)$(PDM)/$P(\nu_\mu \to \nu_\tau)$(MIO) is big,
but since the statistics is poor this ratio is an ambiguous observable   
to establish the difference  between PDM and MIO.

Using Eq.\ (\ref{evtn}) we have estimated the neutrino event rate, 
for the three different hypotheses: no flavor conversion, MIO and PDM.

We addressed a few optimal configurations so that our conclusions may be 
useful in planning real experimental setups.
For each optimal situation, we  have performed our calculation for two 
possible scenarios: (1) a total of $1.6 \times 10^{20}$ $\mu$ decays per year 
with a detector of 10 kton; (2) a total of $2.0 \times 10^{20}$ $\mu$ 
decays per year with a detector of 40 kton. Since the event rate is 
proportional to $E^3_\mu/L^2$ (see Eq.\ (\ref{evtn})), the most conservative 
configuration is  the one with $L_{\text{opt}}= 3096$ km and $E_\mu=10$ GeV.
Our results are summarized in Table \ref{tab4}. It is clear that one can 
distinguish among all the studied hypotheses, even in the most  
modest scenario (1). 

An even more conclusive result can be obtained by looking at the 
energy distribution $dN_\mu/dE_\nu$ (see Eq.\ (\ref{ap13})) for 
charged current $\mu$ events. 
We show, in Fig. \ref{fig4a}, our predictions, for five years of 
data taking, assuming 2.0 $\times 10^{20}$ $\mu$ decays per year, 
$E_\mu=10 $ GeV, $L=3096$ km for null oscillation, the best-fit point 
of the PDM solution to the ANP and four different 
values of $\Delta m^2$ which are consistent with the MIO solution to the ANP.
We see that in the majority of cases a distinct signal will be observed,  
making it possible to establish which of the flavor changing hypotheses 
is the correct one. 

Although we have optimized the setup parameters in order to maximize the 
separation between the two ANP solutions at their best-fit point, we also have
 checked that even if we allow $\gamma_0$ to vary around the best-fit value, 
\ie $0.25 \times 10^{-21} \text{ GeV} \lsim \gamma_0 \lsim  1.1 \times 10^{-21} \text{ GeV}$, 
we will get an energy spectrum which will always increase 
monotonically with the neutrino energy, in this range these curves will all
coinciding up to 4 GeV, after that 
if $\gamma_0<0.6 \times  10^{-21} \text{ GeV}$ the growth will be 
slightly slower than what is shown in Fig. \ref{fig4a}, 
for $\gamma_0=0.6 \times  10^{-21} \text{ GeV}$ and if  
$\gamma_0>0.6 \times  10^{-21} \text{ GeV}$ the curves will be steeper 
approaching the one for no flavor change.  

In Fig.~\ref{fig4b}, we show the 
statistical significance test for 5 years of data taking in the plane 
$\Delta m^2 \times \sin^2 2\theta$ assuming 
the most conservative of the optimal configurations.
From this figure, we see that for most of the MIO parameter space 
the two solutions can be well discriminated.
This conclusion is further confirmed by  Fig.~\ref{fig4c}, 
where the statistical significance test was performed letting $\gamma_0$ 
vary. Here we note that for data compatible with $ 1.5 \times 10^{-3} \text{ eV}^2 \lsim \Delta m^2 \lsim 4.8 \times 10^{-3}$ eV$^2$  a very clear 
separation is possible $\forall \gamma_0$. 
It seems that unless the data prefers lower values 
of $\Delta m^2 \lsim 1.5  \times 10^{-3}$ eV$^2$ or a few isolated 
islands in the plane $\gamma_0 \times \Delta m^2$, 
in which case a clear separation between the two solutions might be 
compromised, decoherence and mass induced oscillation will present 
a very distinct signature of their dynamics. Fortunately, for  
$\Delta m^2 \lsim 1.5  \times 10^{-3}$ eV$^2$ and 
for $\Delta m^2 \gsim 5.0  \times 10^{-3}$ eV$^2$ 
a clear cut between the two solutions can be accomplished respectively 
by the OPERA and the MINOS (medium) experiments, as we have discussed in 
Sec.\ \ref{gsexp} and \ref{minos}.

\section{Conclusions} 
\label{sec:sec5}

We have discussed the perspectives of future experiments of
distinguishing the MIO solution to the ANP from the PDM one. This is 
specially important since it will be very difficult for NOMAD and CHORUS 
to achieve a sensitivity in the $\nu_\mu \to \nu_\tau$ mode to directly 
exclude/confirm the latter solution~\cite{gstz}.

Our study on discriminating these two ANP solutions permitted us to arrive at 
the following general conclusions: K2K and probably MINOS will not be able to 
shed much light on the dynamics which promotes $\nu_\mu \to \nu_\tau$ 
conversion, OPERA and a neutrino factory in a muon collider are 
more suitable for this job.

\begin{figure}
\vglue -3.0cm
\centerline{
\epsfig{file=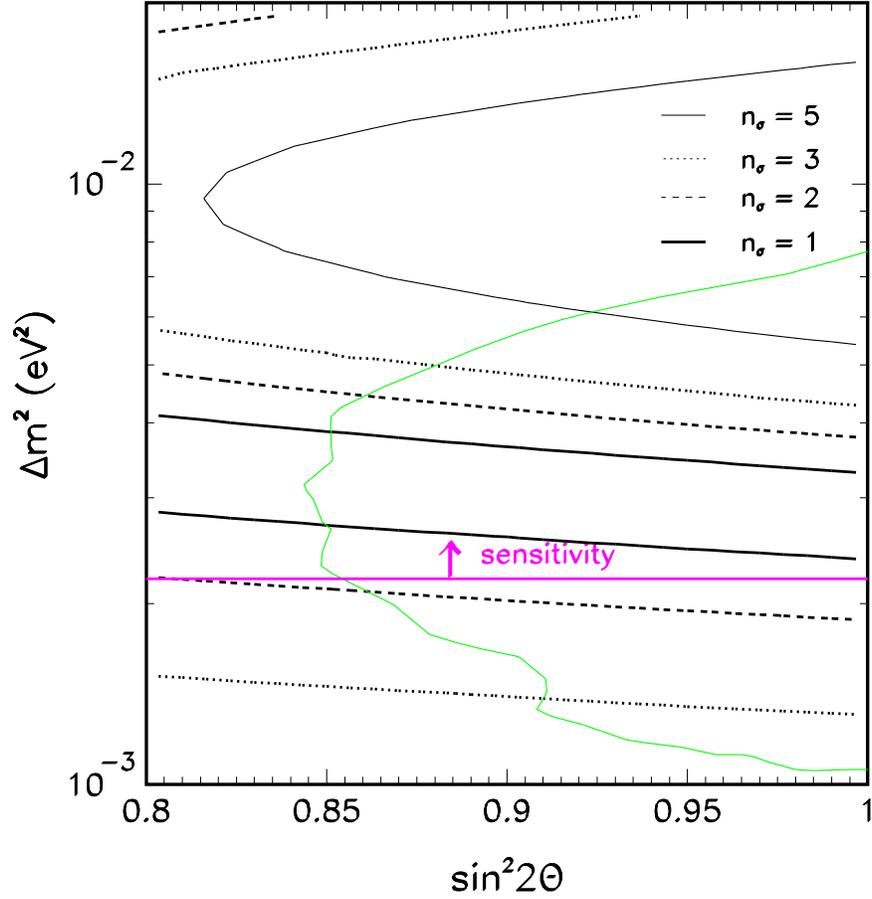,height=20.8cm,width=15cm}}
\vglue -4.5cm
\caption{Regions in the $\sin^2 2 \theta \times \Delta m^2$ plane, 
for $\gamma_0=\gamma_0^{\text{best}}$,  where the number 
of $N_{\text{FD}}$ events expected for PDM and MIO are 
separated by $n_\sigma=1,2,3$ and $5$ for K2K after $10^{20}$ POT.
The inner part of the gray curve is the one allowed at  
99 \% C.L. by the latest SuperK atmospheric neutrino 
data~\protect \cite{sk-atm}. 
The sensitivity of K2K is marked by an horizontal line with an arrow. 
}
\label{fig1b}
\vglue -0.05cm
\end{figure}

\begin{figure}
\vglue -3.0cm
\centerline{
\epsfig{file=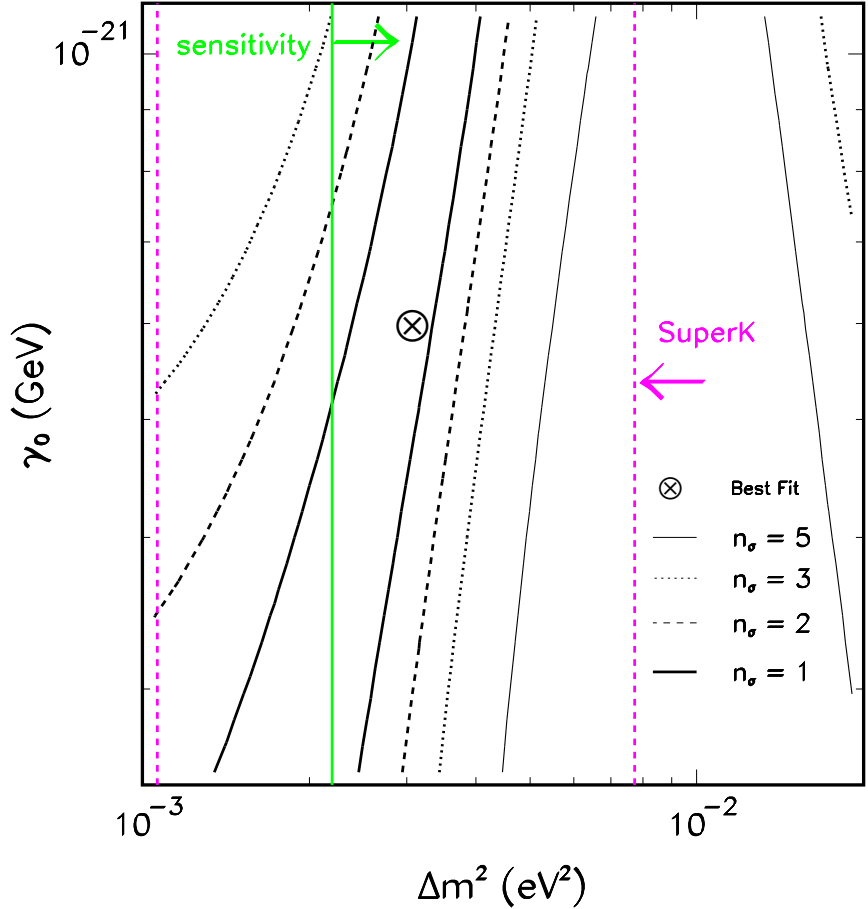,height=20.8cm,width=15cm}}
\vglue -4.5cm
\caption{Regions in the $\gamma_0 \times \Delta m^2$ plane, with 
$\sin^2 2\theta=1$,  
where the number of $N_{\text{FD}}$ events expected 
for PDM and MIO are separated by $n_\sigma=1,2,3$ and $5$  
for K2K after $10^{20}$ POT.
The dotted gray lines mark the region allowed at  
99 \% C.L. by the latest SuperK atmospheric neutrino 
data~\protect \cite{sk-atm} and the cross the best fit values of 
the PDM~\protect \cite{lisi} and the MIO~\protect \cite{sk-atm} solutions 
to the ANP. The start of the sensitivity of K2K is marked by an arrow. 
}
\label{fig1c}
\vglue -0.05cm
\end{figure}

\begin{figure}
\vglue -2.5cm
\centering\leavevmode
\epsfxsize=435pt
\hglue 0.5cm
\epsfbox{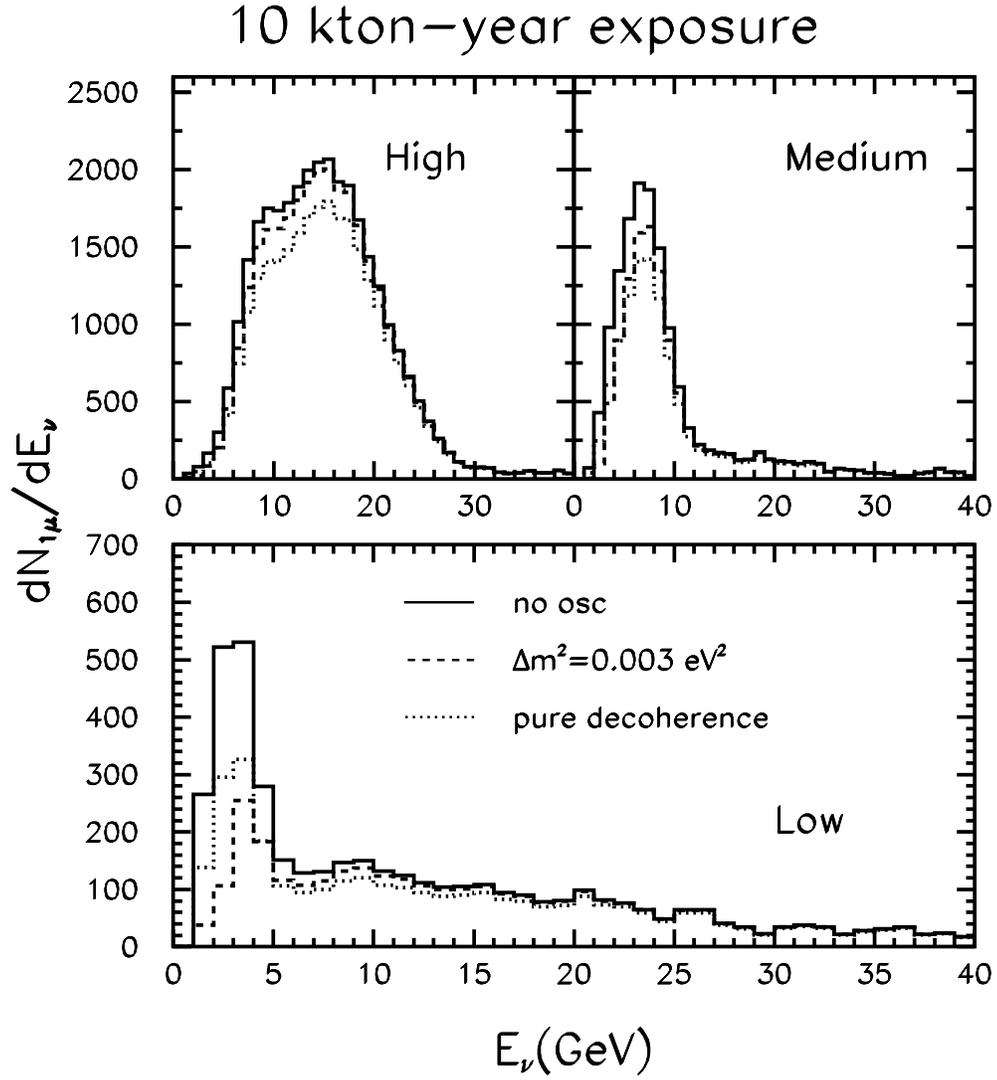}
\vglue -5.0cm
\caption{Spectral distortion expected at MINOS for the best-fit points 
of the vacuum oscillation and of the pure decoherence 
solutions to the ANP for three possible beam configurations.}
\label{fig2a}
\vglue -0.05cm
\end{figure}

\begin{figure}
\vglue -5.0cm
\centerline{
\epsfig{file=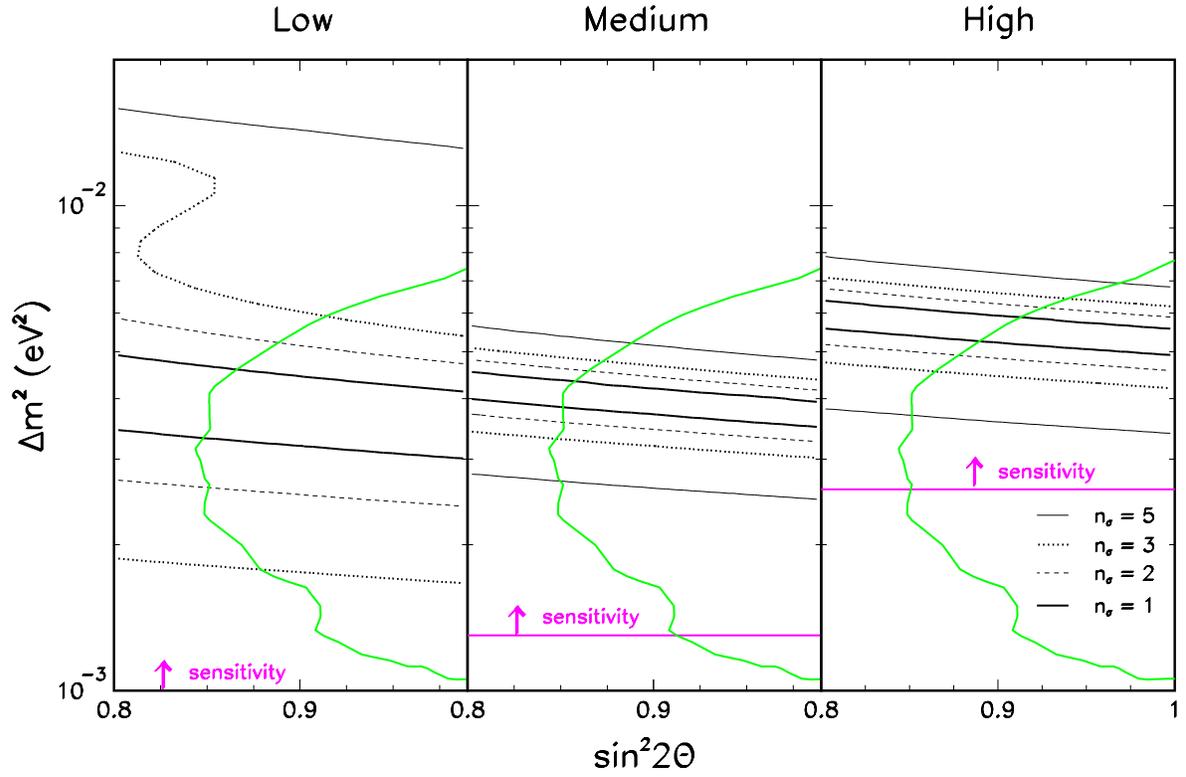,height=10.8cm,width=18.cm}}
\vglue 0.5cm
\caption{ Same as Fig.~\protect \ref{fig1b} but for $N_{1\mu}$ events 
in MINOS after 10 kton-year exposure.
The sensitivity of MINOS is marked by an horizontal line with an arrow. 
}
\label{fig2b}
\vglue -0.05cm
\end{figure}

\begin{figure}
\vglue -5.0cm
\centerline{
\epsfig{file=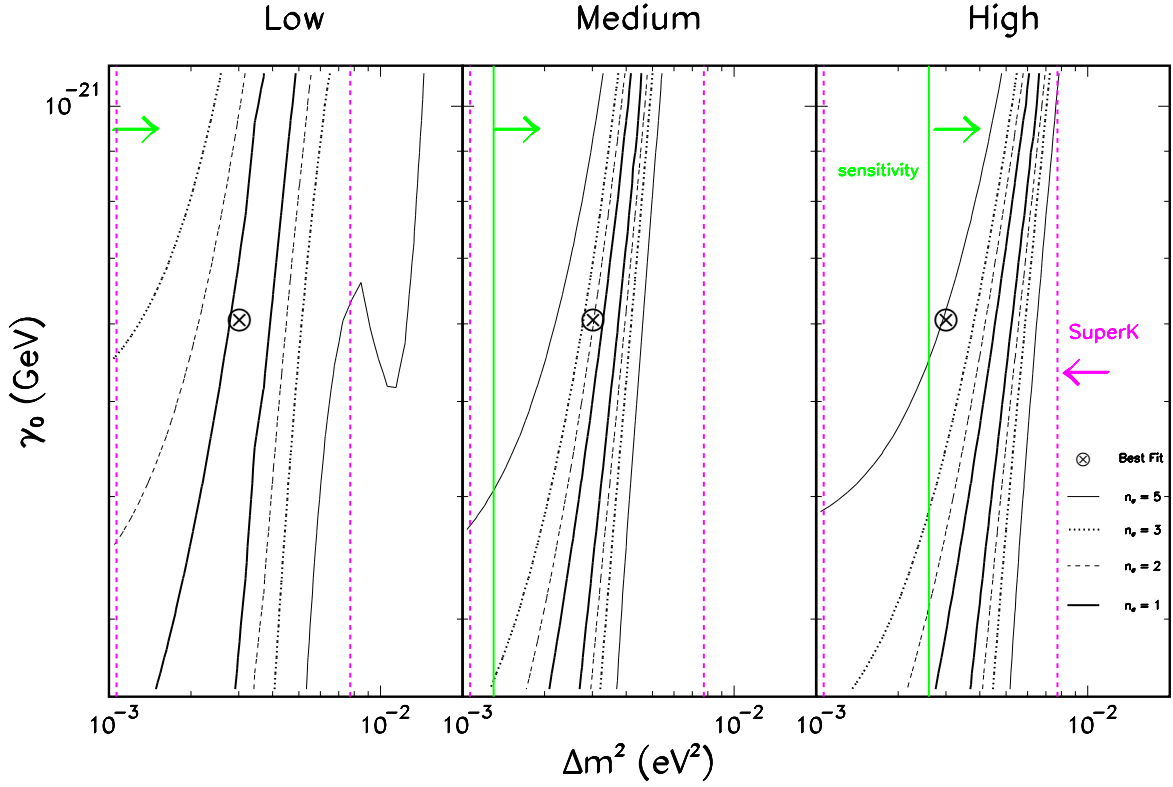,height=10.8cm,width=18cm}}
\vglue 0.5cm
\caption{Same as Fig.~\protect \ref{fig1c} but for $N_{1\mu}$ events 
in MINOS after 10 kton-year exposure. The sensitivity of MINOS is 
marked by a vertical continuous line with an arrow. 
}
\label{fig2c}
\vglue -0.05cm
\end{figure}

\begin{figure}
\vglue -3.5cm
\centering\leavevmode
\epsfxsize=435pt
\hglue -0.5cm
\epsfbox{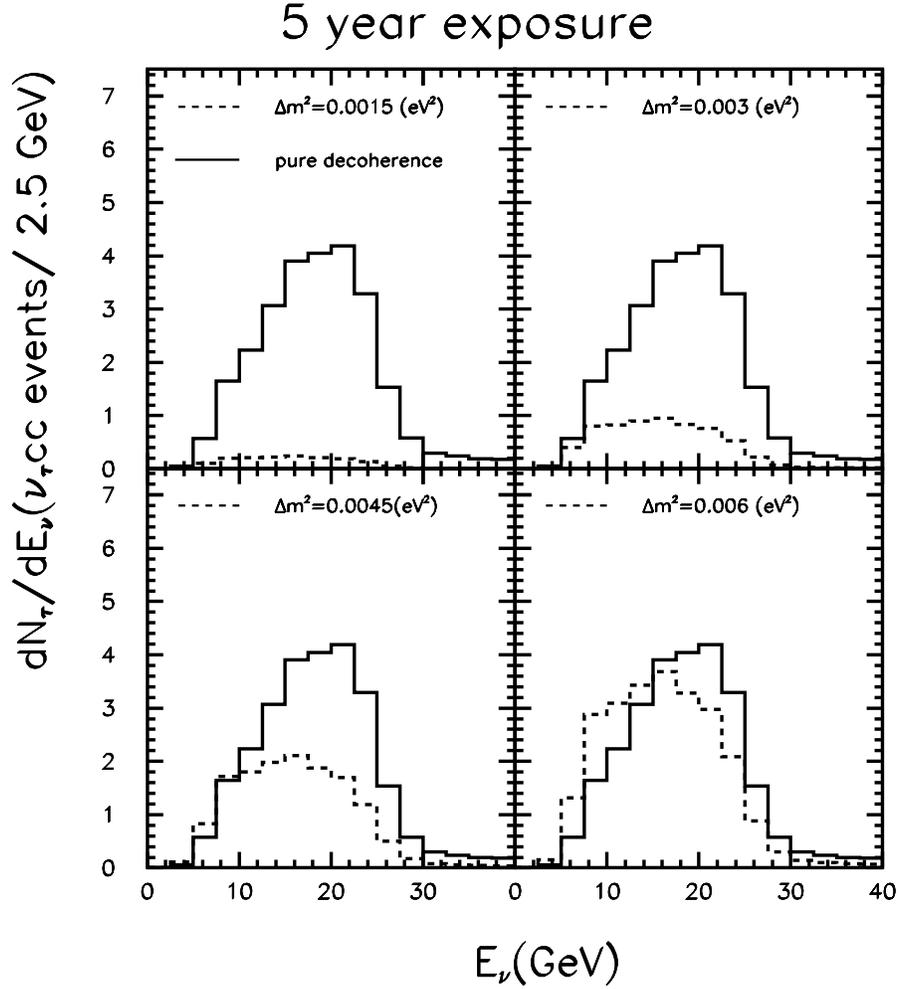}
\vglue -5.0cm
\caption{Spectral distortion expected at OPERA for different values 
of $\Delta m^2$ (dashed lines) as well as for the best-fit 
point of the pure decoherence (continuous line) solution to the ANP, 
for 5 year exposure. We have taken into account the overall efficiency 
of 8.7~\% in accordance to Ref.~\protect \cite{opera}. 
In all cases $\sin^2 2 \theta=1$.}
\label{fig3a}
\vglue -0.05cm
\end{figure}

\begin{figure}
\vglue -3.0cm
\centerline{
\epsfig{file=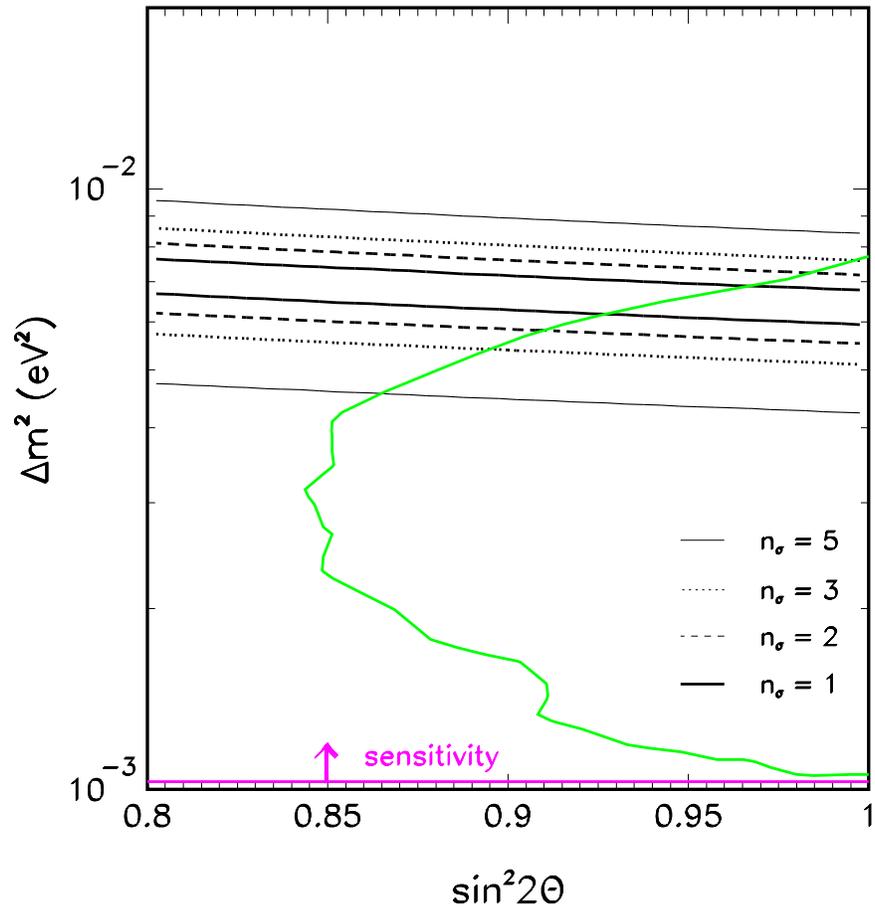,height=20.8cm,width=15cm}}
\vglue -4.5cm
\caption{Same as Fig.~\protect \ref{fig1b} but for $N_\tau$ events 
in OPERA after 5 years.
The sensitivity of OPERA is marked by an horizontal line with an arrow.}
\label{fig3b}
\vglue -0.05cm
\end{figure}

\begin{figure}
\vglue -3.0cm
\centerline{
\epsfig{file=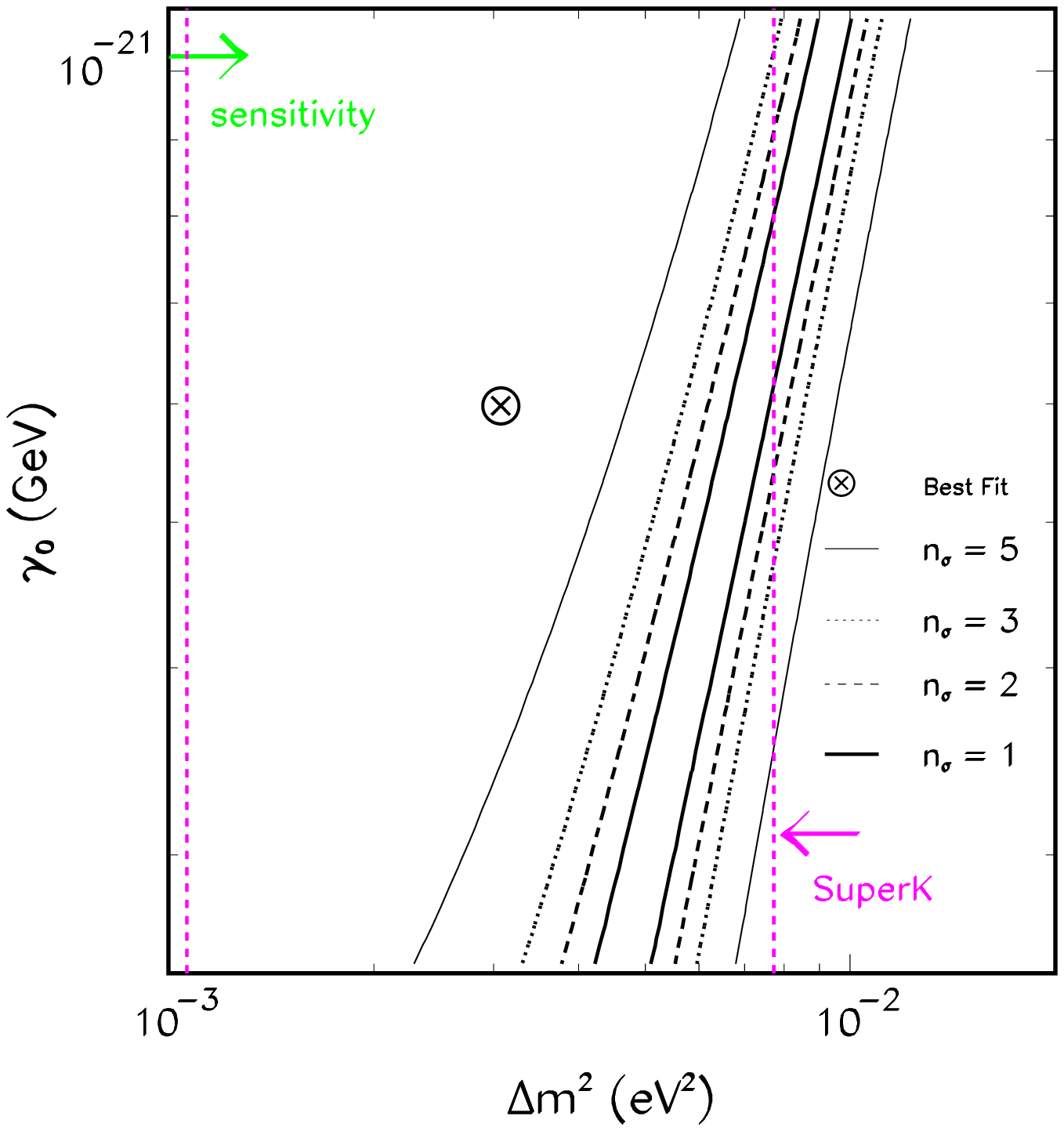,height=20.8cm,width=15cm}}
\vglue -4.5cm
\caption{Same as Fig.~\protect \ref{fig1c} but for $N_\tau$ events 
in OPERA after 5 years. The sensitivity of OPERA is 
marked by an arrow. 
}
\label{fig3c}
\vglue -0.05cm
\end{figure}

\begin{figure}
\vglue -2.5cm
\centering\leavevmode
\epsfxsize=435pt
\hglue 0.5cm
\epsfbox{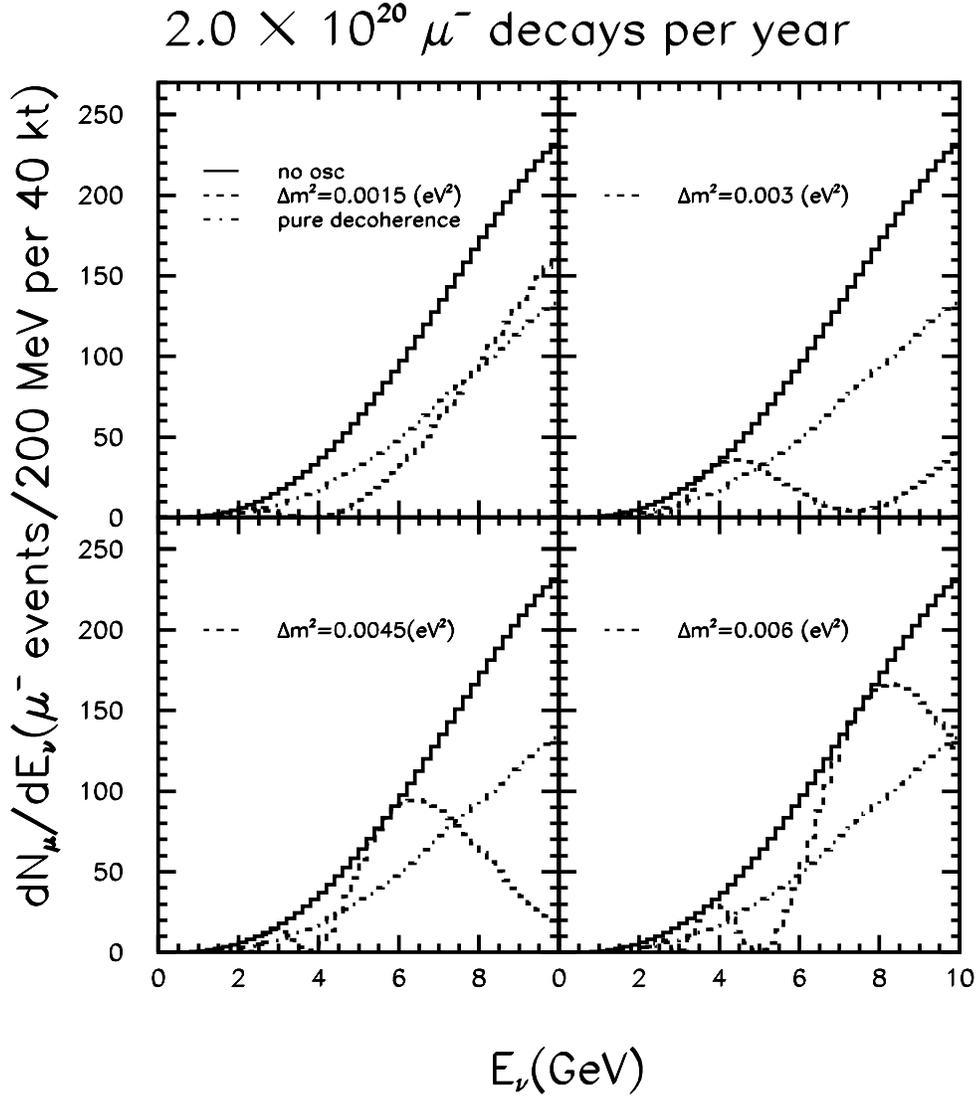}
\vglue -5.0cm
\caption{Spectral distortion expected at a neutrino factory
for $E_\mu=$ 10 GeV, $L=3096$ km and several values of
$\Delta m^2$ as well as for no flavor change (continuous line) and for the 
best-fit point of the pure decoherence (dot-dashed line) solution to the ANP, 
after five years of data taking. In all MIO  cases $\sin^2 2 \theta=1$.}
\label{fig4a}
\vglue -0.05cm
\end{figure}

\begin{figure}
\vglue -3.0cm
\centerline{
\epsfig{file=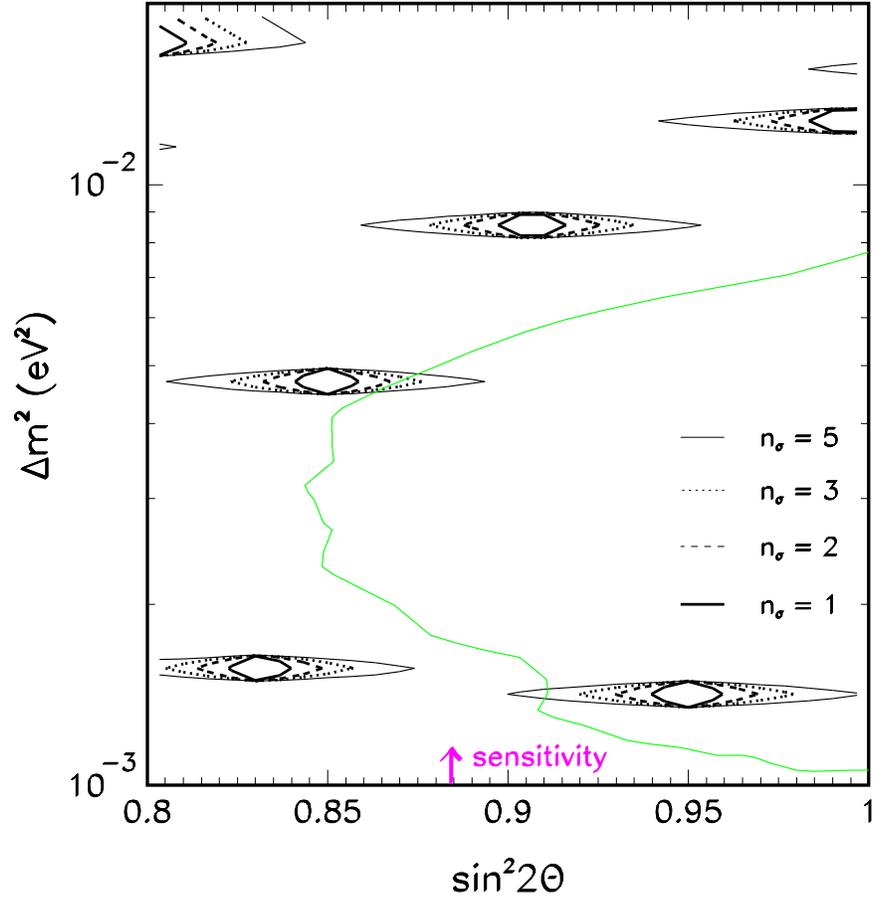,height=20.8cm,width=15cm}}
\vglue -4.5cm
\caption{Same as Fig.~\protect \ref{fig1b} but for $N_{\mu}$ events 
after 5 years of a neutrino factory in the scenario (1) of 
Table \protect \ref{tab4}. The sensitivity of this setup is marked by 
an horizontal line with an arrow.}
\label{fig4b}
\vglue -0.05cm
\end{figure}

\begin{figure}
\vglue -3.0cm
\centerline{
\epsfig{file=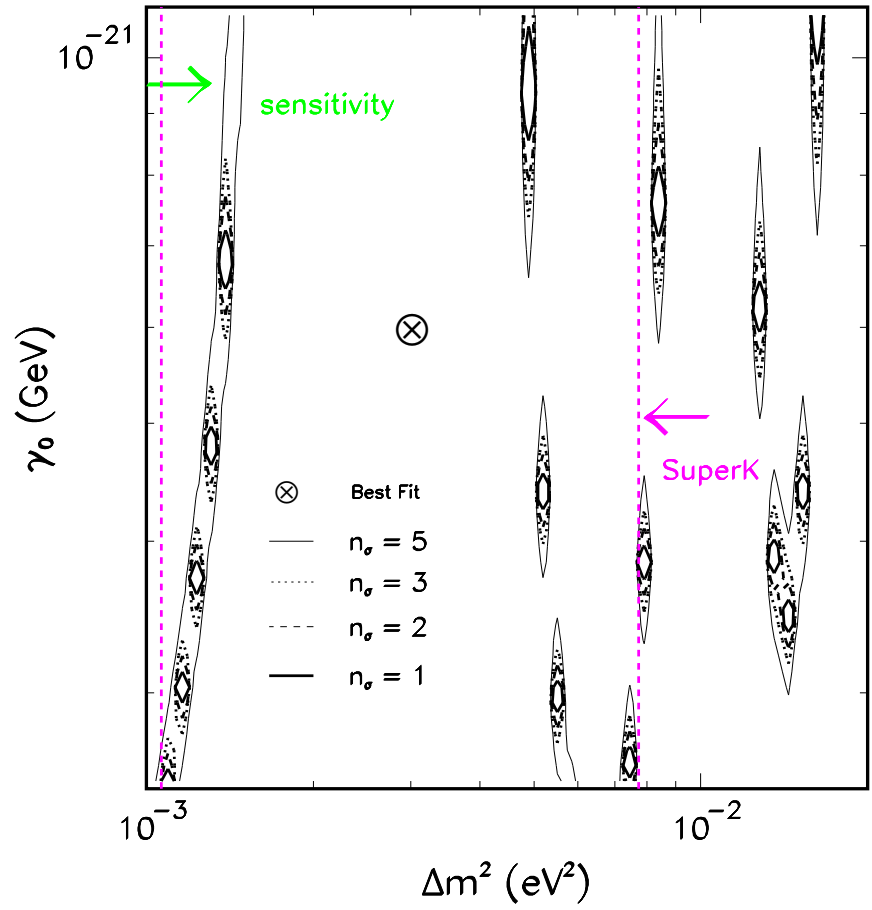,height=20.8cm,width=15cm}}
\vglue -4.5cm
\caption{Same as Fig.~\protect \ref{fig1c} but for $N_\mu$ events 
after 5 years of a neutrino factory in the scenario (1) of 
Table \protect \ref{tab4}. The start of the sensitivity of this setup is 
marked by an arrow. 
}
\label{fig4c}
\vglue -0.05cm
\end{figure}

From the statistical significance tests we have performed, considering 
positive discrimination only if $n_\sigma \gsim 5$, we can say the following. 
K2K cannot  discriminate PDM from MIO 
if the data is compatible with 
$2.2 \times 10^{-3} \text{eV}^2 \lsim \Delta m^2 
\lsim 4.5 \times 10^{-3} \text{eV}^2$, $\forall \gamma_0$.
In fact, this almost covers the entire parameter space 
allowed by the SuperK atmospheric data 
at 99\% C.L., as can be seen in Fig.~\ref{fig1b}.
MINOS, in the low energy beam configuration, cannot separate the 
two solutions if the data is compatible with 
$1.1 \times 10^{-3} \text{eV}^2 \lsim \Delta m^2 \lsim 5.2 \times 10^{-3} \text{eV}^2$, $\forall \gamma_0$; 
in the medium energy beam configuration if 
$3.0 \times 10^{-3} \text{eV}^2 \lsim \Delta m^2 \lsim 4.0 \times 10^{-3} \text{eV}^2$, $\forall \gamma_0$; 
in the high energy beam configuration if $4.8 \times 10^{-3} \text{ eV}^2\lsim \Delta m^2 \lsim 5.1 \times 10^{-3} \text{eV}^2$, $\forall \gamma_0$.
Running with the low energy setup, MINOS will be very similar to K2K, 
and certainly will not be able do discriminate the solutions, see 
Figs.~\ref{fig2b} and \ref{fig2c}. 
For the high energy setup MINOS will be more selective and similar to OPERA. 
OPERA, after 5 years, will only not disentangle PDM from MIO if the data is 
compatible with $5.9 \times 10^{-3} \text{eV}^2 \lsim \Delta m^2 \lsim 6.7 \times 10^{-3} \text{eV}^2$, $\forall \gamma_0$.
This corresponds to the upper corner of the SuperK  
99\% C.L. region, see Fig.~\ref{fig3b}. 
A neutrino factory with $E_\mu=10$ GeV, $L=3096$ km and 
1.6 $\times 10^{20}$ $\mu^-$ decays per year, with a detector 
of 10 kton after 5 years of data taking will still not be able 
to discriminate the two solutions if the data prefers a small island 
in the plane $\sin^2 2\theta \times \Delta m^2$, at 
$\Delta m^2 \sim 1.5 \times 10^{-3}$ eV$^2$ 
and $\sin^2 2\theta \sim 0.95$, see Fig.~\ref{fig4b}. On the other hand, 
if the data is compatible with 
$1.5 \times 10^{-3} \text{eV}^2 \lsim \Delta m^2 \lsim 5 \times 10^{-3} \text{eV}^2$ the separation between PDM and MIO will be extremely 
clear, $\forall \gamma_0$.
This type of facility, it seems, will be the only one able to 
measure spectral distortions and, thereby, directly test 
if neutrino flavor change is indeed an oscillation phenomenon. 

In any case, the combination of the results of all these proposed 
experiments will most certainly unravel the dynamics of neutrino 
conversion in $\nu_\mu \to \nu_\tau$ mode, if only active neutrinos 
exist in nature. 

There is a proposed atmospheric neutrino experiment, MONOLITH~\cite{monolith}, 
which, in principle, could distinguish PDM from MIO through the observation 
of the first oscillation minimum. Here we have only investigated the 
capabilities of accelerator neutrino experiments which we believe have to be 
performed in order to completely ratify the mechanism behind neutrino 
flavor change.

\acknowledgments

We thank GEFAN for valuable discussions 
and useful comments. This work was supported by Conselho Nacional de 
Desenvolvimento Cient\'{\i}fico e Tecnol\'ogico (CNPq) and by Funda\c{c}\~ao 
de Amparo \`a Pesquisa do Estado de S\~ao Paulo (FAPESP).


\newpage


%
%
\vglue -2cm 
\begin{table}
\caption{
Expected number of $\nu_\mu$-events calculated for the K2K Far Detector of 
22.5 kton. In all cases $\sin^2 2\theta=1$.
}
\begin{center}
\begin{tabular}{c|c|c}
Hypothesis & $N_{\text{FD}}$ & Number of POT \\
\hline
No Flavor Change & 40.3 & $2.29 \times 10^{19}$\\
$\Delta m^2=3\times 10^{-3}$ eV$^2$ & 28.3 & \\
$\Delta m^2=5\times 10^{-3}$ eV$^2$ & 18.9 & \\
$\Delta m^2=7\times 10^{-3}$ eV$^2$ & 13.3 & \\
\hline
$\Delta m^2=3\times 10^{-3}$ eV$^2$ & 123 & $1.0 \times 10^{20}$ \\
$\gamma_0=0.6 \times 10^{-21}$ GeV & 122 & \\
\end{tabular}
\label{tab1}
\end{center}
\end{table}

%
%
\vglue -4cm 
\begin{table}
\caption{
Expected ratio $R_{{0\mu}/{1\mu}}$ in MINOS for 10 kton year 
exposure. The error is only statistical. In all cases $\sin^2 2\theta=1$.}
\begin{center}
\begin{tabular}{c|c}
\multicolumn{2}{c}{ Medium Energy Beam } \\ 
\hline
Hypothesis &  $R_{0\mu/1\mu}$ \\
\hline
No Flavor Change &  $0.33 \pm 0.01 $ \\
$\Delta m^2=1.5\times 10^{-3}$ eV$^2$ &  $0.35\pm 0.01$\\
$\Delta m^2=3.0\times 10^{-3}$ eV$^2$ &  $0.41\pm 0.01$\\
$\Delta m^2=4.5\times 10^{-3}$ eV$^2$ &  $0.52\pm 0.01$ \\
$\Delta m^2=6.0\times 10^{-3}$ eV$^2$ &  $0.71 \pm 0.01$ \\
\hline
$\gamma_0=0.6 \times 10^{-21}$ GeV & $0.47 \pm 0.01$ \\
\hline
\hline
\multicolumn{2}{c}{ High Energy Beam } \\ 
\hline
Hypothesis &  $R_{0\mu/1\mu}$ \\
\hline
No Flavor Change &  $0.314 \pm 0.004 $ \\
$\Delta m^2=1.5\times 10^{-3}$ eV$^2$ &  $0.320\pm 0.004$\\
$\Delta m^2=3.0\times 10^{-3}$ eV$^2$ &  $0.340\pm 0.004$\\
$\Delta m^2=4.5\times 10^{-3}$ eV$^2$ &  $0.380\pm 0.005$ \\
$\Delta m^2=6.0\times 10^{-3}$ eV$^2$ &  $0.430 \pm 0.005$ \\
\hline
$\gamma_0=0.6 \times 10^{-21}$ GeV & $0.413 \pm 0.005$ \\
\end{tabular}
\label{tab2}
\end{center}
\end{table}

%
%
\vglue -2cm 
\begin{table}
\caption{
Expected number of $\nu_\tau$-events calculated for OPERA
assuming 5 year exposure, using 8.7\% for the total $\tau$ selection 
efficiency. In all MIO cases $\sin^2 2\theta=1$.}
\begin{center}
\begin{tabular}{c|c}
Hypothesis & $N_{\tau}$ \\
\hline
$\Delta m^2=1.5\times 10^{-3}$ eV$^2$ &  $4$\\
$\Delta m^2=3.0\times 10^{-3}$ eV$^2$ &  $16$\\
$\Delta m^2=4.5\times 10^{-3}$ eV$^2$ &  $36$ \\
$\Delta m^2=6.0\times 10^{-3}$ eV$^2$ &  $62$ \\
\hline
$\gamma_0=0.6 \times 10^{-21}$ GeV & $69$ \\
\hline
\hline
\multicolumn{2}{c}{ Total Background $\sim$ 0.6 events \cite{opera}} \\ 
\end{tabular}
\label{tab3}
\end{center}
\end{table}

%
%
\vglue -3cm 
\begin{table}
\caption{
Expected number of $\mu$ events, $N_\mu$, for a few configurations of 
a neutrino factory under the three studied hypotheses. Scenario (1) 
corresponds to a total of $1.6 \times 10^{20}$ $\mu$ decays per year and a 
10 kton detector, scenario (2) to a total of $2.0 \times 10^{20}$ $\mu$ 
decays per year and a 40 kton detector. The numbers in scenario (1) are 
calculated for one year while in scenario (2) are for five 
years of data taking. The errors given are only statistical.
For MIO we used $\sin^2 2\theta=1$.
}
\begin{center}
\begin{tabular}{c|c|c|c|c|c|c}
Hypothesis & \multicolumn{3}{c|} { $N_\mu$ (1)} &  \multicolumn{3}{c} { $N_\mu$ (2)}   \\
\hline
$L$ (km) & $3096$  & $6192$  & $9289$  & $3096$ & $6192$ & $9289$  \\
$E_\mu$ (GeV)  &  $10$ & $20$ & $30$ & $10$ & $20$ & $30$ \\
\hline
No Flavor Change & $1315 \pm 36 $  & $2580 \pm 50 $ & $3804 \pm 61 $  & $32880 \pm 181 $ & $64501 \pm 253 $ &  $95101 \pm 308 $ \\
\hline
$\Delta m^2 = 3 \times 10^{-3}$ eV$^2$ & $222 \pm 15 $ &$503 \pm 22 $ & $805 \pm 28 $ & $5554\pm 75$ & $12596\pm 112$ & $20141\pm 141$\\
\hline
$\gamma_0=0.6 \times 10^{-21}$ GeV & $716 \pm 27$ & $1439 \pm 37$ &
$2156 \pm 46$ & $ 17918 \pm 134$ & $ 35991 \pm 189$ & $ 53910 \pm 232$ \\ 
\end{tabular}
\label{tab4}
\end{center}
\end{table}

\clearpage


\newpage


%
%

\newpage

%
%

\newpage

%
%

\newpage

%
%

\clearpage

%
%

\clearpage

%
%

\clearpage

%
%

\clearpage 

%
%

\clearpage

%
%

\clearpage

%
%

\newpage 

%
%

\clearpage

%
%


\newpage
\clearpage


\appendix
\section*{}
\label{appendix}

The distribution of ${\nu_\mu}$  in the decay 
${\mu}^- \to e^- + \bar \nu_e + \nu_\mu $ in the muon rest-frame (cm) 
is given by~\cite{geer}

\be
\frac{d^2 N_{\nu_\mu}}{dx\,d{\Omega}_{\text{{cm}}}} =
\frac{1}{4\pi} \left [
h_0(x) + {\cal P}_{\mu} \,
h_1(x) \, \cos{\theta}_{\text{{cm}}} \right]\,,
\label{ap1}
\ee
$x=2 E^{\text{cm}}_{\nu}/m_\mu$, where $E^{\text{cm}}_{\nu}$ denotes the 
neutrino energy, $\theta_{\text{{cm}}}$ is the angle between the neutrino 
momentum vector and the muon spin direction, and ${\cal P}_{\mu}$ is the 
average muon polarization along the beam directions. The functions $h_0$ and 
$h_1$ are given in Table~\ref{tabap}.
  
On applying a Lorentz transformation to boost into the laboratory frame 
(lab), it is found that the neutrino energy distribution at any polar angle 
is just scaled by a relativistic boost factor depending on the angle

\be
E^{\text{{lab}}}_{\nu}= \frac{1}{2} x E^{\text{{lab}}}_{\mu} \left(
1+{\beta}\cos{\theta}_{\text{{cm}}}\right)\,,
\label{ap3}
\ee
and
\be
\sin{\theta}_{\text{{lab}}}=\frac{\sin{\theta}_{\text{{cm}}}}{{\gamma}\left(
1+{\beta}\cos{\theta}_{\text{{cm}}}\right)}\,,
\label{ap4}
\ee
where $\beta$ and $\gamma$ are the usual relativistic factors and we 
have used  ${\gamma}=E^{\text{{lab}}}_{\mu}/m_{\mu}$.

Because in the long-baseline experiments only $\nu_\mu$ emitted in the 
forward direction are relevant to the computed flux, one can 
make the following approximation: $\cos{\theta}_{\text{{cm}}} \simeq 1$, 
$\sin{\theta}_{\text{{cm}}} \simeq {\theta}_{\text{{cm}}}$. Also at high 
energy ${\beta} \simeq 1$. Hence Eq.~(\ref{ap3}) leads to 
$x = E^{\text{{lab}}}_{\nu}/E^{\text{{lab}}}_{\mu}$. Then, we can 
rewrite ${\rm d}{\Omega}_{\text{{cm}}}$ in terms of 
$d{\Omega}_{\text{{lab}}}$ as
\be
 d{\Omega}_{\text{{cm}}}={\gamma}^2\,{\left(
1+{\beta}\right)}^2\,{\rm d}{\Omega}_{\text{{lab}}}\,.
\label{ap5}
\ee

Substituting Eq.~(\ref{ap5}) into Eq.~(\ref{ap1}), we obtain as a function 
of the lab variables  
\be
\frac{d^2 N_{\nu_\mu}}{dx\,d{\Omega}_{\text{{lab}}}} = 
\frac{\left({E^{\text{{lab}}}_{\mu}}\right)^2}{m^2_{\mu}{\pi}} \,
\left [
h_0\left(x\right) \pm {\cal P}_{\mu} \,
h_1\left(x\right)\, \cos{\theta}_{\text{{cm}}}\right]\,,
\label{ap6}
\ee
which for unpolarized muons simplify to
\be
\frac{d^2 N_{\nu_\mu}}{dx\,d{\Omega}_{\text{{lab}}}} =
\frac{\left({E^{\text{{lab}}}_{\mu}}\right)^2}{m^2_{\mu}{\pi}}\,
h_0\left(x\right)\, \; \longrightarrow \;
\frac{d N_{\nu_\mu}}{d{E}_{\nu}^{\text{lab}}} = 4{\pi} 
\frac{E^{\text{{lab}}}_{\mu}}{m^2_{\mu}{\pi}}\,
h_0(x) \,.
\label{ap7}
\ee

The number of expected $\mu$ events from $\nu_\mu$ interactions with the 
detector is given by
\be
N_{\mu} \equiv 
\underbrace{n_{\mu}\,M_d\,10^9\,N_{\text{{A}}}}_{\text{{normalization}}}
\times {\Phi} \times {\sigma} \times {\epsilon}_{\mu}
\label{ap9}
\ee

The $\nu_\mu$ flux $\Phi$ can be  written as
\be
{\Phi} = 
\frac{N_{\nu_\mu}}{4{\pi}L^2}\,\frac{1}{t}\, \; \longrightarrow \;
\frac{d{\Phi}}{d{E}^{\text{{lab}}}_{\nu}} =  \frac{1}{4{\pi}L^2}\,\frac{1}{t}\,
\frac{d N_{\nu_\mu}}{d{E}^{\text{{lab}}}_{\nu}}.
\label{ap10}
\ee

Substituting Eq.~(\ref{ap7}) into Eq.~(\ref{ap10}), we get 
\be
\frac{d{\Phi}}{d{E}^{\text{{lab}}}_{\nu}} =   
\frac{E^{\text{{lab}}}_{\mu}}{m^2_{\mu}{\pi}L^2}\,\frac{1}{t}\,
h_0(x) \,.
\label{ap12}
\ee

\newpage 

Finally, after t-years 
\be
\frac{ d N_{\mu}}{d E^{\text{{lab}}}_{\nu}} =
\frac{n_{\mu}\,M_d\,10^9\,N_{\text{{A}}}}{m^2_{\mu}{\pi}\,L^2}\,
E^{\text{{lab}}}_{\mu}\,h_0(x) \times \sigma \times \epsilon_\mu.
\label{ap13}
\ee

\vglue -3cm
\begin{table}
\caption{Flux functions $h_0(x)$ and $h_1(x)$.}
\begin{center}
\begin{tabular}{|c|c|c|} 
  &                   &            \\              
 &$h_0(x)$ & $h_1(x)$  \\
  &           &           \\ \hline 
  &                                &            \\ 
 & $2x^2(3-2x)$  &  $2x^2(1-2x)$      \\
  &             &    \\  
\end{tabular}
\label{tabap}
\end{center}
\end{table}




\begin{references}


\bibitem{sk98} Super-Kamiokande Collab., Y.\ Fukuda {\it et al.} 
Phys.\ Rev.\ Lett.\ {\bf 81}, 1562 (1998).


\bibitem{soudan2} W.\ Anthony Mann for the Soudan-2 Collaboration, 
hep-ex/0007031.

\bibitem{macro} F.\ Ronga for the MACRO Collaboration, 
Nucl.\ Phys.\ B (Proc. Suppl.) {\bf 87}, 135 (2000). 


\bibitem{k2k} M.\ Sakuda for the K2K Collaboration, talk given at the 
XXX International Conference on High Energy Physics (ICHEP 2000), Osaka, 
Japan, July 27-August 2, 2000, transparencies available 
at http://ichep2000.\-hep.\-sci.\-osaka-u.\-ac.\-jp/\-scan/\-0728/\-pa08/\-sa\-ku\-da/\-index.html. 

\bibitem{mns} Z.\ Maki, M.\ Nakagawa, and S.\ Sakata, Prog.\ Theor.\ Phys.\ 
{\bf 28}, 870 (1962). 



\bibitem{lisi-vep} G.\ L.\ Fogli, E.\ Lisi, A.\ Marrone, G.\ Scioscia, 
Phys.\ Rev.\ D {\bf 60}, 053006 (1999). 


\bibitem{bpl} V.\ Barger, J.\ G.\ Learned, S.\ Pakvasa, T.\ J.\ Weiler,
Phys.\ Rev.\ Lett.\ {\bf 82}, 2640 (1999);
 V.\ Barger, J.\ G.\ Learned, P.\ Lipari, 
M.\ Lusignoli, S.\ Pakvasa, T.\ J.\ Weiler,
Phys.\ Lett.\ B {\bf 462},109 (1999). 



\bibitem{lisi} E.\ Lisi, A.\ Marrone, and D.\ Montanino, Phys.\ Rev.\ Lett.\ 
{\bf 85}, 1166 (2000). 



\bibitem{sk-atm} H.\ Sobel for the Super-Kamiokande Collaboration, 
talk given at the XIX International Conference on 
Neutrino Physics and Astrophysics (Neutrino 2000), Sudbury, Canada, 
June 16-21, 2000, transparencies available at http://nu2000.\-sno.\-lau\-ren\-tian.\-ca/\-H.Sobel/\-index.html.



\bibitem{gptz} A.\ M.\ Gago, O.\ L.\ G.\ Peres, W.\ J.\ C.\ Teves and 
R.\ Zukanovich Funchal, work in preparation.




\bibitem{k2kflux} Y.\ Oyama for the K2K Collaboration, talk given at the YITP workshop on flavor physics, Kyoto, Japan, January 28-30, 1998, hep-ex/9803014.



\bibitem{minos} The Minos Collaboration, {\it ``Neutrino Oscillation
Physics at Fermilab: The NuMI-MINOS Project''}, Fermilab Report 
No. NuMI-L-375 (1998).


\bibitem{opera} OPERA Collaboration, {\em An appearance experiment to search 
for $\nu_\mu \to \nu_\tau$ oscillation in the CNGS beam}, 
CERN/SPSC 2000-028, SPSC/P318,  LNGS P25/2000, July, 2000.



\bibitem{geer}  S.\ Geer, Phys.\ Rev.\ D {\bf 57}, 6989 (1998); 
Phys.\ Rev.\ D {\bf 59}, 039903 (1999).




\bibitem{fb} F.\ Benatti and R.\ Floreanini, JHEP {\bf 0002}, 32 (2000).



\bibitem{ellis1} J.\ Ellis, J.\ S.\ Hagelin, D.\ V.\ Nanopoulos and M.\ 
Srednicki, Nucl.\ Phys.\ B {\bf 241}, 381 (1984).


\bibitem{gstz} A.\ M.\ Gago, E.\ M.\ Santos, W.\ J.\ C.\ Teves and 
R.\ Zukanovich Funchal, Phys.\ Rev.\ D {\bf 63}, 073001 (2001).
\bibitem{kps} H.\ V.\ Klapdor-Kleingrothaus, H.\ P\"as, and U.\ Sarkar, 
Eur.\ Phys.\ J.\ A {\bf 8}, 577 (2000).


\bibitem{pdg00} D.\ E.\ Groom {\it et al.}, European Physical Journal C 
{\bf 15}, 1 (2000).




\bibitem{ratio} Y.\ Oyama for the K2K Collaboration, talk given at the XXXVth 
Rencontres de Moriond ``Electroweak interactions and unified theories'', Les Arc, France, March 11-18, 2000, hep-ex/0004015.

\bibitem{ishida} T.\ Ishida for the K2K Collaboration, hep-ex/0008047.



\bibitem{ngs} A.\ Rubbia for the ICANOE/OPERA Collaboration, 
talk given at the XIX International Conference on Neutrino Physics and 
Astrophysics (Neutrino 2000), Sudbury, Canada, June 16-21, 2000, 
transparencies available at http://nu2000.\-sno.\-lau\-ren\-tian.\-ca/\-A.Rubbia/\-index.html; hep-ex/0008071.




\bibitem{x-sec-k2k} The charged and neutral current cross sections 
were taken from M.\ D.\ Messier, Ph.\ D.\ thesis, Boston University, 1999, 
available at http://hep.bu.edu/$\sim$messier/thesis.




\bibitem{minostec} The MINOS Collaboration, P. Adamson {\it et al.}, 
Fermilab Report No. NuMI-L-337 (1998).


\bibitem{numi} The Minos Collaboration, K.\ R.\ Langenbach and M.\ C.\
Goodman, Fermilab Report  No. NuMI-L-75 (1995).


\bibitem{x-sec} The charged current cross sections for $\nu_\mu$ 
and $\nu_\tau$  can be obtained in form of a table from 
http://www.cern.ch/NGS. 



\bibitem{icanoe} ICANOE Collaboration, {\em ICANOE, 
A proposal for a CERN-GS long baseline and 
atmospheric neutrino oscillation experiment}, INFN/AE-99-17, 
CERN/SPSC 99-25, SPSC/P314, August, 1999, 
see http://pc\-no\-meth4.\-cern.\-ch/\-pu\-bli\-ca\-tions.html.



\bibitem{deruj} A.\ De Rujula, M.\ B.\ Gavela, and P.\ Hernandez, 
Nucl.\ Phys.\ B {\bf 547}, 21 (1999). 
\bibitem{bgw} V.\ Barger, S.\ Geer, and K.\ Whisnant,  
Phys.\ Rev.\ D {\bf 61}, 053004 (2000). 

\bibitem{bgrw} V.\ Barger, S.\ Geer, R.\ Raja, and K.\ Whisnant, 
Phys.\ Rev.\ D {\bf 62}, 013004 (2000). 

\bibitem{flr} M.\ Freund, M.\ Lindner, S.\ T.\ Petcov, A.\ Romanino, Nucl.\ 
Phys.\ B {\bf 578}, 27 (2000).


\bibitem{monolith} K.\ Hoepfner for the MONOLITH Collaboration, 
Nucl.\ Phys.\ B (Proc.\ Suppl.) {\bf 87}, 192 (2000).

\end{references}
\end{document}